\documentclass[journal,12pt,draftcls,onecolumn]{IEEEtran}

\usepackage{cite}
\usepackage{graphicx}

\usepackage[cmex10]{amsmath}
\usepackage{verbatim}
\usepackage{relsize}
\usepackage{amsfonts}
\usepackage{amsthm}
\usepackage{amssymb}
\usepackage{latexsym}
\usepackage{url}
\usepackage{epstopdf}
\usepackage{enumerate}
\usepackage{mathrsfs}
\usepackage{balance}

\newtheorem{theorem}{Theorem}
\newtheorem{lemma}{Lemma}
\newtheorem{definition}{Definition}
\newtheorem{corollary}{Corollary}
\newtheorem{proposition}{Proposition}

\def\myfigwidth{0.7}

\begin{document}

\title{Subblock-Constrained Codes for Real-Time Simultaneous Energy and Information Transfer}

\author{Anshoo~Tandon,~\IEEEmembership{Student~Member,~IEEE,}
~Mehul~Motani,~\IEEEmembership{Member,~IEEE,}
and~Lav~R.\ Varshney,~\IEEEmembership{Senior Member,~IEEE}
\thanks{A. Tandon and M. Motani are with the Department of Electrical and Computer Engineering, National University of Singapore, Singapore 117583 (email: anshoo@nus.edu.sg, motani@nus.edu.sg).}%
\thanks{L.~R.~Varshney is with the Department of Electrical and Computer Engineering and the Coordinated Science Laboratory, University of Illinois at Urbana-Champaign, Urbana, IL 61801 USA (email: varshney@illinois.edu).}%
\thanks{This work was supported in part by the National Research Foundation Singapore under Grant No. NRF-CRP-8-2011-01, and by Systems on Nanoscale Information fabriCs (SONIC), one of the six SRC STARnet Centers, sponsored by MARCO and DARPA.}%
\thanks{Some results in this paper were presented in part at the IEEE SECON 2014 Workshop on Energy Harvesting Communications, June, 2014~\cite{Tandon14_SECON_Workshop}, and at the 2015 International Symposium on Information Theory (ISIT), June, 2015~\cite{Tandon15_ISIT}.}\\
{\vspace{5mm}}
Preprint: Submitted for publication, May 2015.
}

\maketitle
\vspace{-8mm}

\begin{abstract}
Consider an energy-harvesting receiver that uses the same received signal both for decoding information and for harvesting energy, which is employed to power its circuitry. In the scenario where the receiver has limited battery size, a signal with bursty energy content may cause power outage at the receiver since the battery will drain during intervals with low signal energy. In this paper, we consider a discrete memoryless channel and characterize achievable information rates when the energy content in each codeword is regularized by ensuring that sufficient energy is carried within every subblock duration. In particular, we study constant subblock-composition codes (CSCCs) where all subblocks in every codeword have the same fixed composition, and this subblock-composition is chosen to maximize the rate of information transfer while meeting the energy requirement. Compared to constant composition codes (CCCs), we show that CSCCs incur a rate loss and that the error exponent for CSCCs is also related to the error exponent for CCCs by the same rate loss term. We show that CSCC capacity can be improved by allowing different subblocks to have different composition while still meeting the subblock energy constraint. We provide numerical examples highlighting the tradeoff between delivery of sufficient energy to the receiver and achieving high information transfer rates. It is observed that the ability to use energy in real-time imposes less of penalty than the ability to use information in real-time.

\end{abstract}

\section{Introduction}
Although wireless charging of portable electronic devices~\cite{Hui12} and implantable biomedical devices~\cite{Yakovlev12} has attracted the attention of researchers over the last few years, pioneering work on wireless power transfer was conducted over a century ago by Hertz and Tesla~\cite{Brown84}. Similarly, wireless information transfer has a rich history, including works by Popov~\cite{PopovBook}, Bose~\cite{Emerson97}, and Marconi~\cite{Marconi09}. In fact, Marconi's wireless telegraph device, capable of transatlantic radio communication, helped save over 700 lives during the tragic accident of the Titanic in 1912~\cite{Hoolihan12}. However, the first work in an information-theoretic setting on analyzing fundamental tradeoffs between \emph{simultaneous} information and energy transfer is relatively recent~\cite{Lav08}. The study of simultaneous information and energy transfer is relevant for communication from a powered transmitter to an energy-harvesting receiver which uses the same received signal both for decoding information and for harvesting energy. The energy harvested by the receiver is employed to power its circuitry. 

The tradeoff between reliable communication and delivery of energy at the receiver was characterized in~\cite{Lav08} using a general capacity-power function, where transmitted codewords were constrained to have average received energy exceed a threshold. This tradeoff between capacity and energy delivery was extended for frequency-selective channels in~\cite{Grover10}. Since then, there have been numerous extensions of the capacity-power function in various settings~\cite{Popovski13,Ding15,Timotheou15,Zhang15}.
Biomedical applications of wireless energy and information transfer have been proposed through the use of implanted brain-machine interfaces that receive data and energy through inductive coupling~\cite{Harrison07,Yakovlev12,Yilmaz13}.

However, in practical applications such as biomedical, imposing only an average power constraint is not sufficient; we also need to regularize the transferred energy content. This is because a codeword satisfying the average power constraint may still cause outage at the receiver if the energy content in the codeword is bursty, since the receive energy buffer with a relatively small storage capacity may drain during intervals with low signal energy. In order to regularize the energy content in the signal, we herein adopt a \emph{subblock-constrained} approach where codewords are divided into smaller subblocks, and every subblock is constrained to carry sufficient energy exceeding a given threshold. The subblock length and the energy threshold may be chosen to meet the real-time energy requirement at the receiver.

An alternative to the subblock-constraint is the sliding-window constraint, which we do not consider here. Under a sliding-window constraint, each codeword provides sufficient energy within a sliding time window of certain duration. This approach was adopted in~\cite{Barbero11, Fouladgar14}, where the use of runlength codes for simultaneous energy and information transfer was proposed. In~\cite{Tandon14_ITA}, a sliding window constraint was imposed on binary codewords and bounds on the capacity were presented for different binary input channels. Note that the sliding-window constraint is relatively tighter than the subblock-constraint, since  subblock-constraint corresponds to the case where the windows are non-overlapping. 

In this paper, we consider a discrete memoryless channel (DMC) and characterize achievable information rates when each \emph{subblock} is constrained to carry sufficient energy. We assume that corresponding to transmission of each symbol in the input alphabet, the receiver harvests a certain amount of energy as a function of the transmitted symbol. Since different symbols may correspond to different energy levels, the requirement of sufficient energy content within a subblock imposes a constraint on the composition of each subblock. Towards meeting this subblock energy requirement, we introduce the \emph{constant subblock-composition codes} (CSCCs) where all the subblocks in every codeword have the same fixed composition. This subblock-composition, quantifying the fraction of different symbols with each subblock, is chosen to maximize the rate of information transfer while meeting the energy requirement. Note that if $x_1^L$ denotes a given subblock of length $L$, then the composition of $x_1^L$ is the distribution $P_{x_1^L}$ on $\mathcal{X}$ defined by $P_{x_1^L}(x) \triangleq \frac{N(x)}{L}, \ x \in \mathcal{X}$, where $N(x)$ is the number of occurrences of symbol $x$ in subblock $x_1^L$. 

\subsection{Our Contribution}
For meeting the real-time energy requirement at a receiver which uses the received signal to simultaneously harvest energy and decode information, we propose the use of CSCCs (Sec.~\ref{subsec:CSCC_Def}) and establish their capacity as a function of the required energy per symbol (Sec.~\ref{subsec:CSCC_Capacity}).
We show that CSCC capacity can be computed efficiently by exploiting certain symmetry properties (Sec.~\ref{subsec:CSCC_Complexity}) and present bounds on subblock length for avoiding receiver energy outage (Sec.~\ref{subsec:ChoiceOfL}). 

Compared to constant composition codes, we quantify the rate loss incurred due to the additional constraint of restricting all subblocks within codewords to have the same composition (Sec.~\ref{subsec:RatePenalty}). For a given rate of information transfer, we derive a lower bound for the error exponent using CSCC in terms of the error exponent for constant composition codes (Sec.~\ref{subsec:ErrorExponent}).

We show that information rates greater than CSCC capacity can be achieved by allowing different subblocks to have different composition, while still meeting the energy requirement per subblock (Sec.~\ref{Sec:SECC}). 

For enabling real-time information transfer, we consider local subblock decoding where each subblock is decoded independently (Sec.~\ref{Sec:RealTimeInfo}), and compare achievable rates using local subblock decoding with those when all the subblocks within a codeword are jointly decoded. We also provide numerical results highlighting the tradeoff between delivery of sufficient energy to the receiver and achieving high information rates (Sec.~\ref{Sec:NumericalResults}).

\subsection{Related Work}

Codes with different constraints on the codewords have been suggested in the past, depending on the constraints at the transmitter, the properties of the communication channel, or the properties of the storage medium. For digital information storage on magnetic medium \cite{Immink98}, codewords are usually designed to meet the runlength constraint \cite{Immink90} or are optimized for partial response equalization with maximum-likelihood sequence detection (PRML) \cite{Cideciyan92}. The study of information capacity using runlength-limited (RLL) codes on binary symmetric channels (BSC) was carried in \cite{Zehavi88,Shamai90,Jacquet10}. 

A class of binary block codes called \emph{multiply constant-weight codes} (MCWC), where each codeword of length $mn$ is partitioned into $m$ equal parts and has weight $w$ in each part, was explored in \cite{Chee14_MCWC} owing to their potential application in implementation of low-cost authentication methods~\cite{Cherif13}. Note that MCWC, introduced in \cite{Chee14_MCWC} as a generalization of \emph{constant weight codes} \cite{Johnson72}, are themselves a special case of CSCCs with input alphabet size equal to two. When each codeword in an MCWC is arranged as an $m~\times~n$ array and additional weight constraints are imposed on all the columns, the resulting two-dimensional weight constrained codes have potential application in optical storage systems \cite{Ordentlich00} and in power line communications \cite{Chee13}.

Power line communications (PLC) requires the power output to be as constant as possible so that information transfer does not interfere with the primary function of power delivery. One way to achieve this on the PLC channel (which suffers from narrow-band interference, white Gaussian noise, and impulse noise~\cite{Pavlidou03}), is to employ \emph{permutation codes}~\cite{Chu04} where each codeword of length $n$ is a permutation of $n$ different frequencies, with each frequency viewed as an input symbol. Higher rates of information transfer may be achieved using \emph{constant composition codes}~\cite{Chu06} at the cost of local variation in power while ensuring that the power expended is same upon completion of each codeword. When the codeword length is a multiple of the frequency alphabet size, the composition may be chosen such that each frequency occurs equal number times in each codeword~\cite{Chee12}.

The codewords employed by an energy harvesting transmitter are constrained by the instantaneous energy available for transmission. The capacity of these constrained codes over an additive white Gaussian noise (AWGN) channel has been analyzed when the energy storage capability at the transmitter is zero~\cite{Ozel11}, infinite~\cite{Ozel12}, or some finite quantity~\cite{ Dong14,Jog14}. The capacity of an AWGN channel with processing cost at an energy harvesting transmitter was characterized in~\cite{Rajesh14}. The DMC capacity using an energy harvesting transmitter equipped with a finite energy buffer was analyzed in~\cite{Mao13}. A comprehensive summary of the recent contributions in the broad area of energy harvesting wireless communications was provided in~\cite{Ulukus15_Review}.

\section{System Model}

\begin{figure}[t]
\centering
\includegraphics[width=\myfigwidth\textwidth]{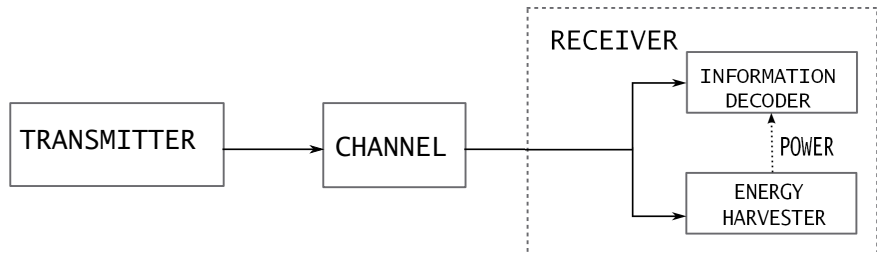}
\caption{Simultaneous information and energy transfer from a transmitter to an energy-harvesting receiver}
\label{Fig:SimultaneousEI}
\end{figure}

Consider communication from a transmitter to a receiver where the receiver uses the received signal both for decoding information as well as for harvesting energy (see Fig.~\ref{Fig:SimultaneousEI}). We model the effective communication channel from the output of a digital modulator at the transmitter to the input to an information decoder at the receiver as a DMC. Note that a DMC is characterized by input alphabet ${\mathcal{X}}$, output alphabet ${\mathcal{Y}}$, and a stochastic matrix $W : \mathcal{X} \to \mathcal{Y}$ with $W = \{W(y|x) : x \in \mathcal{X}, y \in \mathcal{Y}\}$ where the matrix entry $W(y|x)$ is the probability that the output is $y$ when the channel input is $x$.

A DMC is a reasonable communication channel model for simultaneous energy and information transfer. Consider, for instance, the use of a digital modulator at the transmitter which produces symbols from a signal constellation $\mathcal{X} = \{x_1, \ldots, x_{r}\}$. At the receiver, the signal is split for use by the energy harvesting module and the information processing module, respectively. The input to the information decoder at the receiver comprises of one of $s$ quantized values $\mathcal{Y} = \{y_1,\ldots,y_s\}$, fed by a quantizer in the information processing path. For each quantized value $y_i, 1\le i \le s$, and each transmitted symbol $x_j, 1 \le j \le r$, the likelihood $\Pr(y_i|x_j)$ can be computed based on the effective signal path from the transmit modulator to the quantizer at the receiver. The communication channel is thus a DMC with input alphabet $\mathcal{X}$, output alphabet $\mathcal{Y}$, and channel transition probabilities $\Pr(y_i|x_j)$.

In practice, the effective channels seen by the information decoder and the energy harvester may be different due to their respective pre-processing stages. A simple time-sharing approach to transmitting energy and information simultaneously was suggested in~\cite{LavThesis} via interleaving of energy signal and information-bearing signal. In~\cite{Zhou13}, practical architectures for simultaneous information and energy reception were defined: an ``integrated" receiver architecture has shared radio frequency chains between the energy harvester and the information decoder, whereas a ``separated" architecture has different chains.

In our work, we assume a generic receiver architecture where the received signal is split between the energy harvesting path and the information processing path with a static power splitting ratio. The effective communication channel seen by the decoder in the information processing path is modeled as a DMC. We let $b(x)$ denote the energy harvested by the harvester after the signal split at the receiver, when $x \in \mathcal{X}$ is transmitted. Thus, $b$ is a map from the input alphabet $\mathcal{X}$ to the set of non-negative real numbers, and higher energy is carried by symbols having higher $b$-value. This map is assumed to be time-invariant, and reflects the scenario where the statistical nature of the effective communication channel is due to the noise in the receiver circuitry, which does not affect the harvested energy. The quantification of $b$ abstracts the specific implementation of a chosen receiver architecture, which in turn helps to abstract the problem of the code design for simultaneous energy and information transfer from implementation details.

\begin{figure}[t]
\centering
\includegraphics[width=\myfigwidth\textwidth]{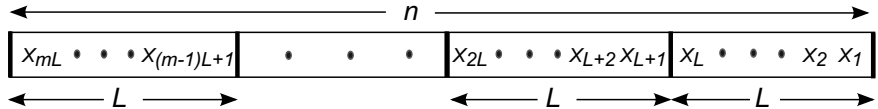}
\caption{Transmitted codeword partitioned into subblocks of length $L$.}
\label{Fig:codeword}
\end{figure}

In order to meet the real-time energy requirement at the receiver, we partition the transmitted codeword into equal-sized subblocks (see Fig.~\ref{Fig:codeword}) and require that transmitted symbols be chosen such that the expected harvested energy in each subblock exceeds a given threshold. This threshold is a function of the energy consumption by the receiver circuitry including the information decoder. We will denote the subblock length by $L$ and assume that the codeword length, denoted $n$, is a multiple of $L$. If a transmitted codeword is denoted $(X_1, X_2, \ldots, X_n)$, then the constraint on sufficient energy within each subblock can be expressed as
\begin{equation}
\frac{1}{L} \sum_{i=1}^{L} b\left( X_{(j-1)L + i} \right) \ge B , \ \ j = 1, 2, \ldots , m
\label{eq:SubblockEnergyConstraint}
\end{equation}
where $j$ is the subblock index, $B$ denotes the required energy per symbol at the receiver, and $m$ is the number of subblocks in a codeword. 
The choice of the subblock length $L$ depends on the energy storage capacity at the receiver; a small energy buffer generally requires relatively small value of $L$ to prevent energy outage at the receiver. 

The subblock energy constraint given by \eqref{eq:SubblockEnergyConstraint} becomes trivial if $b(x)$ is same for all $x \in \mathcal{X}$ (for instance, when the transmitted symbols belong to a phase-shift-keying constellation). However, the constraint is non-trivial when $b$-values are not constant (for instance, using on-off keying) and threshold $B$ satisfies
\begin{equation}
b_{\min} < B < b_{\max} ,
\label{eq:EnergyThreshold}
\end{equation}
where 
\begin{equation}
b_{\min} = \min_{x \in \mathcal{X}} b(x), \ \ \ \ \ b_{\max} = \max_{x \in \mathcal{X}} b(x).
\end{equation}
In the rest of the paper we assume \eqref{eq:EnergyThreshold} is satisfied, unless otherwise stated. 

For a given subblock $j$ within a codeword, if $N(x)$ denotes the number of occurrences of symbol $x$ in the $j$th subblock, then \eqref{eq:SubblockEnergyConstraint} can alternately be expressed as
\begin{equation}
\sum_{x \in \mathcal{X}} b(x) \frac{N(x)}{L} \ge B .
\label{eq:SubblockEnergyConstraint2}
\end{equation}
Note that $N(x)/L$ denotes the fraction of time when symbol $x$ appears in the subblock. We now introduce constant subblock-composition codes which are a nice way to meet the subblock energy constraint.

\section{Constant Subblock-Composition Codes}
\subsection{Motivation and Definition}
\label{subsec:CSCC_Def}
We have seen that for a given subblock, the energy constraint given by \eqref{eq:SubblockEnergyConstraint} can equivalently be expressed as \eqref{eq:SubblockEnergyConstraint2} and this constraint is satisfied provided the fraction of time each symbol appears in the subblock is chosen appropriately. This observation motivates the use of codes where the composition of each subblock in all codewords is constant and is chosen such that \eqref{eq:SubblockEnergyConstraint2} is satisfied. 
A \emph{constant subblock-composition code} (CSCC) is one in which all codewords are partitioned into equal-sized subblocks and each subblock (in all codewords) has the same type $P$. The subblock type $P$ in CSCC is chosen to satisfy the subblock energy constraint
\begin{equation}
\mathbb{E}_{P}\left[b(X)\right] \triangleq \sum_{x \in \mathcal{X}} b(x) P(x) = \sum_{x \in \mathcal{X}} b(x) \frac{N(x)}{L} \ge B .
\label{eq:SubblockEnergyConstraint3}
\end{equation}

\subsection{Capacity using CSCC}
\label{subsec:CSCC_Capacity}
Let $\mathcal{P}_L$ denote the set of all compositions for input sequences of length $L$. For a given type $P \in \mathcal{P}_L$, the set of sequences in $\mathcal{X}^L$ with composition $P$ is denoted by $\mathcal{T}_{P}^{L}$ and is called the \emph{type class} or \emph{composition class} of $P$. In a CSCC with subblock-composition $P$, every subblock in a codeword may be viewed as an element of $\mathcal{T}_{P}^{L}$. 

In order to compute the capacity of a CSCC on a DMC, we may view the $L$ uses of the original channel as a single use of the induced \emph{vector channel} having input alphabet $\mathcal{T}_{P}^{L}$ and output alphabet $\mathcal{Y}^L$. Since the underlying channel is memoryless, the transition probabilities for a pair of input and output vectors is the product of the corresponding transition probabilities of the underlying channel. If we let $x_1^L = x_1 \ldots x_L$ and $y_1^L = y_1 \ldots y_L$ be given input and output vectors with $x_i \in \mathcal{X}$ and $y_i \in \mathcal{Y}$, respectively, then the transition probabilities for the induced vector channel are:
\begin{equation}
W^L(y_1^L | x_1^L) = \prod_{i=1}^{L} W(y_i|x_i) .
\label{eq:VectorChannel}
\end{equation}

Since each subblock in a codeword may be chosen independently, the capacity using CSCC with subblock-composition $P$, denoted $C_{CSCC}^{L}(P)$, is equal to $1/L$ times the capacity of the induced vector channel with input alphabet $\mathcal{T}_{P}^{L}$, output alphabet $\mathcal{Y}^L$, and transition probabilities given by \eqref{eq:VectorChannel}. Thus if we denote $X_1^L = X_1\ldots X_L$ and $Y_1^L = Y_1\ldots Y_L$, then
\begin{align}
C_{CSCC}^{L}(P) &= \max_{X_1^L \in \mathcal{T}_{P}^{L}} \frac{I(X_1^L ; Y_1^L)}{L} \label{eq:CSCC_P_Capacity0} \\
&= \max_{X_1^L \in \mathcal{T}_{P}^{L}}  \left( \frac{H(Y_1^L)}{L} - \frac{H(Y_1^L | X_1^L)}{L} \right) \\
&= \max_{X_1^L \in \mathcal{T}_{P}^{L}} \left( \frac{H(Y_1^L)}{L} - \frac{ \sum_{i=1}^{L} H(Y_i | X_i)}{L} \right)
\label{eq:CSCC_P_Capacity}
\end{align}
where the last equality follows from the memoryless property of the channel. The maximization in \eqref{eq:CSCC_P_Capacity0} is over the distribution of input vector in $\mathcal{T}_{P}^{L}$. We will show that the maximum is achieved when the input vectors $X_1^L$ are uniformly distributed over $\mathcal{T}_{P}^{L}$.

\begin{theorem}
The capacity of the induced vector-channel using CSCC with fixed subblock-composition $P$ is obtained via a uniform distribution of the input vectors in $\mathcal{T}_{P}^{L}$.
\label{th:UniformDistInputVectors}
\end{theorem}
\begin{IEEEproof}
See Appendix~\ref{app:UniformDistInputVectors}.
\end{IEEEproof}

If we define the set of distributions
\begin{equation}
\Gamma_{B}^{L} \triangleq \{ P \in \mathcal{P}_L : \ \mathbb{E}_P[b(X)] \ge B \} ,
\label{eq:ProbabilityDistributionSet}
\end{equation}
then the capacity using CSCC with subblock energy constraint \eqref{eq:SubblockEnergyConstraint}, denoted $C_{CSCC}^L(B)$, is defined as
\begin{equation}
C_{CSCC}^L(B) = \max_{P \in \Gamma_{B}^{L}} C_{CSCC}^{L}(P)
\label{eq:CSCC_Capacity}
\end{equation}

\subsection{Computing CSCC Capacity}
\label{subsec:CSCC_Complexity}
By Theorem~\ref{th:UniformDistInputVectors}, the maximum is achieved in \eqref{eq:CSCC_P_Capacity} when $X_1^L$ is uniformly distributed over $\mathcal{T}_{P}^{L}$. The computation of the capacity expression with increasing subblock length $L$ seems challenging since the input and output alphabet size for the induced vector channel grows exponentially with $L$. However, we will show that the computational complexity of the CSCC capacity expression can be reduced using the following observations.

First note that the probability distribution for the output vector in the induced vector channel is given by
\begin{equation}
P_{Y_1^L}(y_1^L) = \frac{1}{|\mathcal{T}_{P}^{L}|} \sum_{x_1^L \in \mathcal{T}_{P}^{L}} W^L(y_1^L |x_1^L) ,
\label{eq:OutputProbDist}
\end{equation}
since the input vectors are uniformly distributed over $\mathcal{T}_{P}^{L}$. If $\tilde{y}_1^L$ is another output vector having the same composition as $y_1^L$, then we have $P_{Y_1^L}(y_1^L) = P_{Y_1^L}(\tilde{y}_1^L)$. This is because the columns $W^{L} (y_1^L | \cdot)$ and $W^{L} (\tilde{y}_1^L | \cdot)$ of the vector channel transition matrix are permutations of each other (see Appendix~\ref{app:UniformDistInputVectors}). Thus output vectors having the same composition have equal probability. However, even though the input vectors are uniformly distributed, the output vectors in general are \emph{not} uniformly distributed. Also, since the symbols within an input vector $x_1^L \in \mathcal{T}_{P}^{L}$ are not independent, in general we have $P_{Y_1^L}(y_1^L) \neq \prod_{i=1}^L P_Y(y_i)$, where $P_Y(y)$ denotes the probability of output scalar symbol $y$. 

Let $\mathcal{Q}_L$ denote the set of all compositions for output sequences of length $L$. When $X_1^L$ is uniformly distributed over $\mathcal{T}_{P}^{L}$, the $H(Y_1^L)$ term in \eqref{eq:CSCC_P_Capacity} can be expressed as
\begin{align}
H(Y_1^L) &= - \sum_{y_1^L \in \mathcal{Y}^L} P_{Y_1^L}(y_1^L) \log P_{Y_1^L}(y_1^L) \\
&= - \sum_{Q \in \mathcal{Q}_L} \sum_{y_1^L \in \mathcal{T}_{Q}^{L}} P_{Y_1^L}(y_1^L) \log P_{Y_1^L}(y_1^L) \\
&= \sum_{Q \in \mathcal{Q}_L} |\mathcal{T}_{Q}^{L}| P_{Y_1^L}(y_1^L) \log \frac{1}{P_{Y_1^L}(y_1^L)} ,
\label{eq:JointEntropy}
\end{align}
where the last equality follows because $P_{Y_1^L}(y_1^L)$ is same for all $y_1^L \in \mathcal{T}_{Q}^{L}$. Note that we choose only one representative vector $y_1^L$ from each type class $\mathcal{T}_{Q}^{L}$ in the last equality.

Secondly, the following proposition shows that the $H(Y_i | X_i)$ term in \eqref{eq:CSCC_P_Capacity} is same for all $1 \le i \le L$, since the corresponding joint probabilities $P_{XY} (X_i = x , Y_i = y)$ are equal. 
\begin{proposition}
\label{prop:PairwiseProb}
For a random input vector $X_1^L$ uniformly distributed over $\mathcal{T}_{P}^{L}$  with corresponding output vector $Y_1^L$, the pairwise probability $P_{XY} (X_i = x , Y_i = y)$, for $1 \le i \le L$, satisfies
\begin{equation}
P_{XY} (X_i = x , Y_i = y) = \frac{N(x)}{L} W(y|x) = P(x) W(y|x) .
\label{eq:PairwiseProb}
\end{equation}
\end{proposition}
\begin{IEEEproof}
Since 
\begin{equation}
P_{XY} (X_i = x , Y_i = y) \ = \ \mathrm{Pr}(X_i = x) W(y|x) ,
\end{equation}
the claim will be proved if we show $\mathrm{Pr}(X_i = x) = N(x)/L$ for all $1 \le i \le L$. As $X_1^L$ is uniformly distributed over $\mathcal{T}_{P}^{L}$, the $\mathrm{Pr}(X_i = x)$ is equal to the ratio of the number of input vectors with $x$ at index $i$ to the total number of vectors in $\mathcal{T}_{P}^{L}$. Since 
\begin{equation}
|\mathcal{T}_{P}^{L}| = \frac{L!}{\displaystyle \prod_{x \in \mathcal{X}} N(x)!} ,
\label{eq:TotalInputVector}
\end{equation}
and the number of sequences in $\mathcal{T}_{P}^{L}$ with $x$ at index $i$ is
\begin{equation}
\frac{(L-1)!}{\displaystyle \left(N(x)-1\right)! \prod_{\tilde{x} \neq x} N(\tilde{x})!} ,
\label{eq:PartialVectors}
\end{equation}
the ratio of the quantities given by \eqref{eq:PartialVectors} and \eqref{eq:TotalInputVector} is equal to $\mathrm{Pr}(X_i = x) = N(x)/L$.
\end{IEEEproof}

The next proposition gives a computationally efficient expression for CSCC capacity.
\begin{proposition}
The CSCC capacity, $C_{CSCC}^L(B)$, is given by
\begin{equation}
\max_{P \in \Gamma_{B}^{L}} \frac{1}{L} \sum_{Q \in \mathcal{Q}_L} |\mathcal{T}_{Q}^{L}| \  P_{Y_1^L}(y_1^L) \log \frac{1}{P_{Y_1^L}(y_1^L)} - H(Y|X) ,
\label{eq:CSCC_Capacity2}
\end{equation}
where only one representative output vector $y_1^L$ is chosen from every type class $\mathcal{T}_{Q}^{L}$, $P_{Y_1^L}(y_1^L)$ is given by \eqref{eq:OutputProbDist}, and
$H(Y|X)$ is evaluated using the joint pairwise probability distribution given by \eqref{eq:PairwiseProb}.
\label{prop:CSCC_Capacity2}
\end{proposition}
\begin{IEEEproof}
Use \eqref{eq:CSCC_P_Capacity} and \eqref{eq:CSCC_Capacity} to express $C_{CSCC}^L(B)$. From Thm.~\ref{th:UniformDistInputVectors}, a uniform distribution over $\mathcal{T}_{P}^{L}$ achieves capacity, and hence the entropy term $H(Y_1^L)$ in \eqref{eq:CSCC_P_Capacity} can be computed using \eqref{eq:JointEntropy}. The claim in Prop.~\ref{prop:CSCC_Capacity2} follows by further noting that the $H(Y_i | X_i)$ term in \eqref{eq:CSCC_P_Capacity} is the same for all $1 \le i \le L$, which can be evaluated using the joint pairwise distribution in \eqref{eq:PairwiseProb}.
\end{IEEEproof}

\subsection{Choice of Subblock Length $L$}
\label{subsec:ChoiceOfL}
In this subsection, we derive bounds on subblock length $L$ (as a function of the energy storage capacity at the receiver) which will ensure that the receiver never runs out of energy when the subblock-composition $P$ is chosen to satisfy \eqref{eq:SubblockEnergyConstraint3}. It will be seen that a large energy storage capacity allows for larger values of $L$ and hence results in higher rates of information transfer.

The energy storage capacity at the receiver is denoted $E_{max}$ and we assume that the receiver requires $B$ units of energy per received symbol for its processing. Let $E(i)$ denote the level of the energy buffer at the receiver at the completion of $i-1$ uses of the channel. The energy update equation, for $i=1, 2, \ldots \,$, is given by
\begin{equation}
E(i+1) = \min\left(E_{max}, |E(i) + b(X_i) - B|^{+} \right) ,
\end{equation}
where $X_i$ is the symbol transmitted in the $i$th channel use, and $|z|^{+} \triangleq \max(z,0)$. 

We say that an \emph{outage} occurs during $i$th channel use if 
\begin{equation}
E(i) + b(X_i) < B ,
\end{equation}
while an \emph{overflow} event occurs if
\begin{equation}
E(i) + b(X_i) - B > E_{max} \ .
\end{equation}
We partition the input alphabet as $\mathcal{X} = \mathcal{X}_{\triangleleft} \cup \mathcal{X}_{\triangleright}$, where
\begin{align}
\mathcal{X}_{\triangleleft} &= \{ x \in \mathcal{X} \, | \, b(x) < B \} \ , \\
\mathcal{X}_{\triangleright} &= \{ x \in \mathcal{X} \, | \, b(x) \ge B \} \ .
\end{align} 
For CSCC with subblock-composition $P \in \Gamma_{B}^{L}$, we define
\begin{equation}
G = \sum_{x \in \mathcal{X}_{\triangleleft}} L P(x) \left(B - b(x)\right) \ ,
\label{eq:G_def}
\end{equation}
where $G$ will be used to characterizes some useful properties of the energy update process.

\begin{lemma}
\label{lemma:NoOutage}
The energy update process satisfies the following properties for CSCC with subblock-composition $P \in \Gamma_{B}^{L}$:
\begin{enumerate}[(a)]
\item If there is no energy outage or overflow during the reception of the first subblock, then $E(L+1) \ge E(1)$.
\item If $E(1) \ge G$, then there is no energy outage during the reception of the first subblock.
\item If $E(1) \ge G$ and $E_{max} \ge 2 G$, then $E(L+1) \ge G$.
\end{enumerate}
\end{lemma}
\begin{IEEEproof}
If there is no energy outage or overflow, then the total energy harvested during the reception of the first subblock is $\sum_{x \in \mathcal{X}} L P(x) b(x)$, while the total energy consumed is $L B$ and claim $(a)$ follows since $P$ satisfies \eqref{eq:SubblockEnergyConstraint3}.

Let $X_i$ denote the transmitted symbol in the $i$th  channel use, $I = \{1,2,\ldots,L\}$, and $I_{<} = \{ i \in I | X_i \in \mathcal{X}_{\triangleleft} \}$. For $i \in I$, the level in the energy buffer decreases during the $i$th  channel use if and only if $i \in I_{<}$, and the corresponding decrease in energy level is $B-b(X_i)$. Since the subblock has composition $P$, the sum of energy decrements over the reception of the first subblock is $\sum_{i \in I_{<}} B-b(X_i) = G$, and claim $(b)$ follows.

For proving claim $(c)$, we note that the condition $E(1) \ge G$ implies that there is no energy outage during the reception of the first subblock (using claim $(b)$). Further, if there is no overflow then $E(L+1) \ge E(1) \ge G$ (using claim $(a)$). In case there is energy overflow in the $i$th  channel use for any $i \in I$, we have $E(i+1) = E_{max} \ge 2 G$, and thus $E(L+1) \ge E(i+1) - G \ge G$.
\end{IEEEproof}

Lemma~\ref{lemma:NoOutage} is useful in proving the following theorem which gives a necessary and sufficient condition on subblock length in order to avoid outage. 

\begin{theorem}
\label{th:BoundOnL}
A necessary and sufficient condition on $L$ for avoiding energy outage during the reception of CSCC codewords, with subblock-composition $P$ satisfying \eqref{eq:SubblockEnergyConstraint3}, is 
\begin{equation}
L \le \frac{E_{max}}{\sum_{x \in \mathcal{X}_{\triangleleft}} 2 P(x) \left(B - b(x)\right)} ,
\label{eq:BoundOnL}
\end{equation}
with $E(1) \ge G$.
\end{theorem}

\begin{IEEEproof}
See Appendix~\ref{app:BoundOnL}.
\end{IEEEproof}

The initial condition on energy level, $E(1) \ge G$, may be ensured by transmitting a preamble, consisting of symbols with high energy content, before the transmission of codewords. This preamble has bounded length and hence does not affect the channel capacity.

\section{Comparing CSCC with Constant Composition Codes}
\label{Sec:CSCC_vs_CCC}
\subsection{Rate Comparison}
\label{subsec:RatePenalty}
Similar to subblock-composition, a codeword composition represents the fraction of times each input symbol occurs in a codeword and a constant composition code (CCC) is one in which all codewords have the same composition. 
Note that a CSCC with subblock-composition $P$ may also be viewed as a CCC with codeword composition $P$, since all the subblocks in CSCC have the same composition. In general for CCC, although all codewords have the same composition, different subblocks within a codeword may have different compositions. Hence CCCs are richer than CSCCs in terms of choice of symbols within each subblock. CCCs were first analyzed by Fano \cite{FanoBook61} and shown to be sufficient to achieve capacity for any discrete memoryless channel.

Let $C_{CCC}(P)$ denote the maximum achievable rate using CCC with codeword composition $P$. For $P \in \Gamma_{B}^{L}$ (refer \eqref{eq:ProbabilityDistributionSet}), a CCC with \emph{codeword} composition $P$ will ensure that the average received energy per symbol in a codeword is at least $B$. However, it may violate the constraint on providing sufficient energy to the receiver \emph{within every subblock duration}. For a CCC, we have \cite{FanoBook61}
\begin{equation}
C_{CCC}(P) = I(X;Y) = H(X) - H(X|Y) .
\label{eq:CCC_Capacity}
\end{equation}

We are interested in quantifying the information rate penalty incurred by using CSCC compared to CCC, given by $C_{CCC}(P) - C_{CSCC}^{L}(P)$. This information rate penalty is the price we pay for meeting the real-time energy requirement within every subblock duration, compared to the less constrained energy requirement per codeword. Although the rate penalty can be numerically computed by explicit computation of $C_{CCC}(P)$ and $C_{CSCC}^{L}(P)$, the numerical approach has the limitation that the computation complexity of $C_{CSCC}^{L}(P)$ increases with an increase in subblock $L$.

In CSCC, since a transmitted subblock $X_1^L$ is uniformly distributed over $\mathcal{T}_{P}^{L}$, we have \cite[p.~26]{CsiszarBook_IIed}
\begin{equation}
H(X_1^L) = \log |\mathcal{T}_{P}^{L}| = L H(P) - L \, r(L,P) , \label{eq:TypeClassSize}
\end{equation}
where $r(L,P)$ denotes a function of $L$ and $P$ given as
\begin{equation}
r(L,P) = \frac{s(P)-1}{2L} \log(2\pi L) + \frac{1}{2L} \sum_{a:P(a)>0} \log P(a) + \frac{\vartheta(L,P)}{12 L \ln 2} s(P) ,
\label{eq:RateLoss}
\end{equation}
with $s(P)$ denoting the number of elements $x \in \mathcal{X}$ with $P(x)>0$, and $\vartheta(L,P)$ is a real number between zero and one which is chosen so that \eqref{eq:TypeClassSize} is satisfied. 

We now present simple analytical bounds for this rate penalty. The following theorem shows that the rate penalty by using CSCC, relative to CCC, is bounded by $r(L,P)$.

\begin{theorem}
The rate penalty is bounded as
\begin{equation}
0 \le C_{CCC}(P) - C_{CSCC}^{L}(P) \le r(L,P) .
\label{eq:RateLossBound0}
\end{equation}
Further, there exist channels for which the rate penalty meets the upper or lower bound in \eqref{eq:RateLossBound0} with equality.
\end{theorem}
\begin{IEEEproof}
When $X_1^L$ is uniformly distributed over $\mathcal{T}_{P}^{L}$,
\begin{align}
C_{CSCC}^{L}(P) &= \frac{1}{L} \left[ H(X_1^L) - H(X_1^L | Y_1^L) \right] \\
&\overset{(a)}{=}  H(P) - r(L,P) - \frac{1}{L} \sum_{i=1}^{L} H \left(X_i |Y_1^L, X_1^{i-1}\right) \nonumber \\
&\overset{(b)}{\ge}  H(P) - r(L,P) - \frac{1}{L} \sum_{i=1}^{L} H(X_i | Y_i) \nonumber \\
&\overset{(c)}{=} H(P) - r(L,P) - H(X|Y) \nonumber \\
&\overset{(d)}{=} C_{CCC}(P) - r(L,P) ,
\label{eq:Diff_CSCC_CCC}
\end{align}
where $X_1^{i-1}$ denotes $X_1,\ldots,X_{i-1}$, $(a)$ follows from \eqref{eq:TypeClassSize} and chain rule for entropy, $(b)$ follows since conditioning only reduces entropy, $(c)$ follows from \eqref{eq:PairwiseProb}, and $(d)$ follows from \eqref{eq:CCC_Capacity}. Now, \eqref{eq:RateLossBound0} follows from \eqref{eq:Diff_CSCC_CCC}. Explicit channels can be constructed which meet the bounds in \eqref{eq:RateLossBound0}.
\begin{itemize}
\item $C_{CCC}(P) = C_{CSCC}^{L}(P) = 0$ for a binary symmetric channel (BSC) with crossover probability equal to $0.5$. 
\item For a noiseless channel, we have $C_{CCC}(P) - C_{CSCC}^{L}(P) = r(L,P)$ due to equality in $(b)$ as $\sum_{i=1}^{L} H(X_i | Y_1^L, X_1,\ldots,X_{i-1}) = \sum_{i=1}^{L} H(X_i | Y_i) = 0$.
\end{itemize}
\end{IEEEproof}

\begin{corollary}
\begin{equation}
\lim_{L \to \infty} C_{CSCC}^L(P) = C_{CCC}(P)
\label{eq:C_CSCC_L_infty}
\end{equation}
\end{corollary}
\begin{IEEEproof}
Note that for a fixed $P$, the value of $r(L,P)$ as a function of $L$ is non-negative and falls roughly as $\log(L)/L$ and thus tends to zero as $L \to \infty$. Thus \eqref{eq:C_CSCC_L_infty} follows by taking the limit $L \to \infty$ in \eqref{eq:RateLossBound0}.
\end{IEEEproof}

\emph{Remark}: For a fixed subblock length $L$, the CSCC capacity can be achieved by making the number of subblocks in a codeword arbitrarily large and performing joint decoding over all the subblocks. However, when the number of subblocks in a codeword are kept constant and the subblock length is increased without bounds, then achievable rates using CSCC tend to CCC capacity. In particular, when there is only one subblock in a codeword, then the CSCC code is same as a CCC code whose capacity can be achieved by making $L$ arbitrarily large.

The upper bound \eqref{eq:RateLoss} on the rate penalty given by $r(L,P)$ is independent of the underlying channel. In general, given a communication channel, the bounds on rate penalty can be further improved. Consider, for example, a BSC with crossover probability $p_0$ where $0 < p_0 < 0.5$. For this channel, the upper bound can be tightened using Thm.~\ref{th:RateLossBound}. 
We first define a binary operator~$\star$ and a function $h$, respectively, as
\begin{align}
a \star b &\triangleq a(1-b) + (1-a)b . \\
h(x) &\triangleq -x \log x - (1-x) \log (1-x) . \label{eq:h_def}
\end{align}
We employ the above definitions to state the following theorem on bounding the rate penalty for a BSC.
\begin{theorem}
\label{th:RateLossBound}
For a BSC with crossover probability $0 < p_0 < 0.5$, input distribution denoted by $P(0) = \Pr(X=0), \ P(1) = \Pr(X=1)$, and $0 < \gamma = \min(P(0),P(1)) \le 0.5$ we have, 
\begin{equation}
0 < C_{CCC}(P) - C_{CSCC}^{L}(P) \le h(p_0 \star \gamma) - h(p_0 \star \alpha) < r(L,P) ,
\label{eq:RateLossBound_BSC}
\end{equation}
where $\alpha$ is chosen such that 
\begin{equation}
h(\alpha) = h(\gamma) - r(L,P) , \ \ 0 \le \alpha < 0.5 \ .
\label{eq:AlphaDef}
\end{equation}
\end{theorem}
\begin{IEEEproof}
See Appendix~\ref{app:RateLossBound}.
\end{IEEEproof}
The proof of Theorem~\ref{th:RateLossBound} uses Mrs. Gerber's Lemma (MGL) \cite{Wyner73}.
Using an extension \cite{Chayat89} of MGL, the upper bound on the rate penalty can similarly be improved for general memoryless binary-input symmetric-output channels. In particular, we have the following theorem for the binary erasure channel (BEC).
\begin{theorem}
\label{th:RateLossBound_BEC}
For a BEC with erasure probability $\epsilon > 0$, 
\begin{equation}
C_{CCC}(P) - C_{CSCC}^{L}(P) \le (1-\epsilon) r(L,P) < r(L,P)
\label{eq:RateLossBound_BEC}
\end{equation}
\end{theorem}
\begin{IEEEproof}
See Appendix~\ref{app:RateLossBound_BEC}.
\end{IEEEproof}
For memoryless asymmetric binary-input, binary-output channels, an alternate upper bound on the rate penalty (other than \eqref{eq:RateLossBound0}) may be obtained using the equality of the channel \emph{characteristic function} and the \emph{gerbator} \cite{Ahlswede77}. As an example, we have the following theorem for the $Z$-channel.
\begin{theorem}
\label{th:RateLossBound_Z}
For a $Z$-channel with $\gamma = \Pr(X=1)$, and $p_0 = \Pr(1\rightarrow 0)$, we have
\begin{equation}
C_{CCC}(P) - C_{CSCC}^{L}(P) \le h\left(\gamma(1-p_0)\right) - h\left(\alpha(1-p_0)\right),
\label{eq:RateLossBound_Z}
\end{equation}
where $h(\cdot)$ is given by \eqref{eq:h_def}, and $\alpha$ is chosen such that
\begin{equation}
h(\alpha) = h(\gamma) - r(L,P) , \ \ 0 \le \alpha < 0.5 \ .
\end{equation}
\end{theorem}
\begin{IEEEproof}
See Appendix~\ref{app:RateLossBound_Z}.
\end{IEEEproof}
The rate penalty bound given by \eqref{eq:RateLossBound_Z} may sometimes be worse than the bound in \eqref{eq:RateLossBound0}, depending on $\gamma$ and $p_0$. In general, the rate penalty for the $Z$-channel can be upper bounded by $\min\left( r(L,P), \, h\left(\gamma(1-p_0)\right) - h\left(\alpha(1-p_0)\right) \right)$.

\subsection{Error Exponent Comparison}
\label{subsec:ErrorExponent}
In this subsection, we discuss the error exponent using CSCC and show that it can be bounded as a function of the (computationally simpler) error exponent for CCC.

We now present some definitions and notations which will be used in this subsection. For a pair of random variables $(X,Y)$ with $P_X = P$, and conditional probability distribution $P_{Y|X} = W$, we will write $H(Y|X)$ as $H(W|P)$, $I(X;Y)$ as $I(P,W)$, and the distribution of $Y$ as $PW$. Thus we have
\begin{align}
PW(y) &\triangleq \sum_{x \in \mathcal{X}} P(x) W(y|x), \ \ y \in \mathcal{Y} \\
H(W|P) &\triangleq \sum_{x \in \mathcal{X}} P(x) H\left(W(\cdot | x)\right) \\
I(P,W) &\triangleq H(PW) - H(W|P) \ .
\end{align}
The \emph{informational divergence} of distributions $P$ and $Q$ is denoted as
\begin{equation}
D(P||Q) \triangleq \sum_{x \in \mathcal{X}} P(x) \log \frac{P(x)}{Q(x)} \ .
\end{equation}
The \emph{conditional informational divergence} of stochastic matrices $V : \mathcal{X} \to \mathcal{Y}$ and $W : \mathcal{X} \to \mathcal{Y}$ with respect to distribution $P$ on $X$ is denoted as 
\begin{equation}
D(V||W | P) \triangleq \sum_{x \in \mathcal{X}} P(x) D\left(V(\cdot |x) || W(\cdot | x)\right) \ .
\end{equation}

For CCCs with codeword composition $P$ and information rate $R > 0$, the \emph{sphere packing exponent function} \cite{CsiszarBook_IIed} of DMC $W$ is given by 
\begin{equation}
E_{sp}(R,P,W) \triangleq \min_{V : I(P,V) \le R} D(V || W | P) \ ,
\label{eq:SP_EE}
\end{equation}
with $V$ ranging over all channels $V: \mathcal{X} \to \mathcal{Y}$, and represents an upper bound on the error exponent using best possible codes. For fixed $P$ and $W$, the function $E_{sp}(R,P,W)$ is a convex function of $R > 0$ (which follows from convexity of $D(V || W | P)$ and $I(P,V)$ as a function of $V$), positive for $R < I(P,W)$ and zero otherwise.

The \emph{random coding exponent function} \cite{CsiszarBook_IIed} of channel $W$ for CCCs with codeword composition $P$ and information rate $R > 0$ is denoted by $E_r(R,P,W)$ and represents a lower bound on achievable error exponent. It is related to $E_{sp}(R,P,W)$ as
\begin{equation}
E_r(R,P,W) = \begin{cases}
E_{sp}(R,P,W) , & \mathrm{if \ } R \ge \hat{R} \\
E_{sp}(\hat{R},P,W) + \hat{R}-R , & \mathrm{if \ } 0 < R < \hat{R} ,
\end{cases}
\label{eq:R_EE}
\end{equation}
where $\hat{R}$ is the smallest $R$ at which the convex curve $E_{sp}(R,P,W)$ meets its supporting line of slope $-1$. 

The structure of $V$ which achieves the minimum in \eqref{eq:SP_EE} for $R < I(P,W)$ is given by the following lemma. For $R \ge I(P,W)$, the minimum in \eqref{eq:SP_EE} is equal to zero which is obtained by choosing $V=W$.  
\begin{lemma}
\label{lemma:StructureOfV}
For $R < I(P,W)$, the stochastic matrix $V:\mathcal{X} \to \mathcal{Y}$ which minimizes $D(V||W|P)$ subject to $I(P,V) \le R$ is given by
\begin{equation}
V(y|x) = \frac{W(y|x)^{1-s} PV(y)^s}{\sum_{\tilde{y} \in \mathcal{Y}} W(\tilde{y}|x)^{1-s} PV(\tilde{y})^s} ,
\label{eq:OptV}
\end{equation}
where $PV(y)$ satisfies the set of simultaneous equations
\begin{equation}
PV(y) = \sum_{x \in \mathcal{X}} P(x) V(y|x) = \sum_{x \in \mathcal{X}} \frac{P(x) W(y|x)^{1-s} PV(y)^s}{\displaystyle \sum_{\tilde{y} \in \mathcal{Y}} W(\tilde{y}|x)^{1-s} PV(\tilde{y})^s} ,
\label{eq:OptPV}
\end{equation}
and $s \in [0,1]$ is chosen such that $I(P,V) = R$.
\end{lemma}

\begin{IEEEproof}
See Appendix~\ref{app:StructureOfV}.
\end{IEEEproof}
We remark that the random coding exponent function for a DMC was stated by Fano \cite{FanoBook61} using the distributions $V(y|x)$ and $PV(y)$, given by \eqref{eq:OptV} and \eqref{eq:OptPV}, respectively, and were referred to as \emph{tilted probability distributions}. However, the explicit statement of Lemma~\ref{lemma:StructureOfV} seems not to have appeared in the literature before.

The following theorem uses Shannon's random coding argument to bound the probability of error for CSCC with subblock-composition $P$ on a DMC. It also applies Lemma~\ref{lemma:StructureOfV} to compactly express the error probability in terms of the sphere packing exponent function.

\begin{theorem}
\label{th:CSCC_ErrorProb}
There exists a CSCC with subblock length $L$, subblock-composition $P$, and codeword length $n$, transmitting information at rate $R>0$ on DMC $W$, for which the maximum probability of error is upper bounded as
\begin{equation}
P_e < \begin{cases}
2 \exp\left(-n E_{sp}(R',P,W)\right), & \mathrm{if \ } R' \ge \hat{R} \\
\exp\left(-n \left(E_{sp}(\hat{R},P,W)+\hat{R}-R'\right)\right), & \mathrm{if \ }  R' < \hat{R} ,
\end{cases}
\label{eq:Pe_CSCC}
\end{equation}
where $R' = R + r(L,P)$ and $\hat{R}$ is the smallest $R'$ at which the convex curve $E_{sp}(R',P,W)$ meets its supporting line of slope $-1$.
\end{theorem}

\begin{IEEEproof}
See Appendix~\ref{app:CSCC_ErrorProb}.
\end{IEEEproof}
The following corollary is immediate.
\begin{corollary}
The error exponent for CSCC with subblock length $L$, subblock-composition $P$, information rate $R>0$ on DMC $W$, is lower bounded by
\begin{equation}
E_r\left(R+r(L,P),P,W\right) .
\end{equation}
\end{corollary}
Thus the bound on the error exponent for CSCC is related to the error exponent for CCC by the same term, $r(L,P)$, as the bound for the rate penalty \eqref{eq:RateLossBound0}.

\section{Beyond Constant Subblock Composition Codes}
\label{Sec:SECC}
In a CSCC, every subblock within any codeword has the same composition, and this composition is chosen to meet the subblock energy constraint \eqref{eq:SubblockEnergyConstraint3}. The capacity using CSCC (given by \eqref{eq:CSCC_Capacity}) is achieved by choosing that subblock-composition in $\Gamma_{B}^{L}$ (given by \eqref{eq:ProbabilityDistributionSet}) which maximizes the information rate. We will see that rates greater than $C_{CSCC}^L(B)$ can be achieved while still meeting the subblock energy constraint \eqref{eq:SubblockEnergyConstraint}.

We first review known results when constraints are placed on the entire codeword (with no subblock constraints) \cite{Lav08, CsiszarBook_IIed}. Let $X_1^n = (X_1, X_2, \ldots, X_n)$ denote any codeword of length $n$. If we impose the average energy constraint on codewords, 
\begin{equation}
\frac{1}{n} \sum_{i=1}^n b(X_i) \ge B , 
\label{eq:CodewordEnergyConstraint}
\end{equation}
then the channel capacity with this constraint is \cite{Lav08, CsiszarBook_IIed}
\begin{equation}
\max_{P_X : \, \mathbb{E}_{P_X}[b(X)] \ge B} I(X;Y) .
\label{eq:ConstrainedCapacity}
\end{equation}
Information rates arbitrarily close to this capacity can be achieved by making the codeword length sufficiently large. Moreover, if $P_X^*$ is an input distribution which maximizes \eqref{eq:ConstrainedCapacity}, then this capacity can be achieved by a sequence of CCCs with codeword composition tending to $P_X^*$ \cite{FanoBook61, CsiszarBook_IIed}. Thus, if $C_{CCC}(B)$ denotes the capacity using CCC when the average energy per symbol is constrained to be at least $B$, then
\begin{align}
C_{CCC}(B) &= \max_{P : \, \mathbb{E}_{P}[b(X)] \ge B} C_{CCC}(P) \label{eq:CCC_B_Capacity} \\
&= \max_{P_X : \, \mathbb{E}_{P_X}[b(X)] \ge B} I(X;Y) .
\end{align}
Thus the capacity with codeword constraints can be achieved by restricting the codewords to have a fixed composition. This is possible because for a given transmission rate, the codebook size increases exponentially with codeword length $n$ while the number of different types of sequences only increase polynomially with $n$. 

We will now show that contrary to the case with codeword constraints, when the constraints are applied to fixed sized subblocks then information rates can, in general, be increased by \emph{not} restricting the subblocks to have a fixed composition. Towards this, we define a \emph{subblock energy-constrained code} (SECC) as a code which satisfies the subblock energy constraint given by \eqref{eq:SubblockEnergyConstraint}. Since all subblocks in SECC satisfy \eqref{eq:SubblockEnergyConstraint}, the composition of each subblock belongs to the set $\Gamma_{B}^{L}$.

Let $C_{SECC}^{L}(B)$ denote the capacity using SECC with subblock length $L$ and average energy per symbol at least $B$. Similar to CSCC, the $L$ uses of the channel in case of SECC induce a vector channel with input alphabet 
\begin{equation}
\mathcal{A} = \underset{P \in \Gamma_{B}^{L}}{\bigcup} \mathcal{T}_{P}^{L} \ ,
\label{eq:SECC_InputAlphabet}
\end{equation}
output alphabet $\mathcal{Y}^L$, and channel transition probabilities given by \eqref{eq:VectorChannel}. Since each subblock may be chosen independently,
\begin{equation}
C_{SECC}^{L}(B) = \max_{X_1^L \in \mathcal{A}} \frac{I(X_1^L ; Y_1^L)}{L} \ ,
\label{eq:SECC_Capacity} 
\end{equation}
where the maximization is over the probability distribution of input vectors in $\mathcal{A}$. For a \emph{noiseless} $q$-ary channel ($\mathcal{X}=\mathcal{Y}=\{0,1,\ldots,q-1\}, \ W(i|i)=1, i \in \mathcal{X}$), it is easy to check that SECC capacity is achieved by the uniform distribution of $X_1^L$ over $\mathcal{A}$. Thus for the noiseless channel, we have $C_{SECC}^L(B) = \log |\mathcal{A}| / L$.

For CSCC, the induced vector channel was symmetric (irrespective of the underlying (scalar) DMC being symmetric or not), and hence the capacity was achieved with a uniform distribution over the input alphabet. In contrast, in case of SECC the induced vector channel need not be symmetric even when the underlying DMC is symmetric. This is formalized in the following theorem which is proved by providing a counterexample.
\begin{theorem}
\label{th:SECC_NotUniform}
Uniform distribution of $X_1^L$ over $\mathcal{A}$ may not achieve SECC capacity even when the underlying DMC is symmetric.
\end{theorem}
\begin{IEEEproof}
See Appendix~\ref{app:SECC_NotUniform}.
\end{IEEEproof}
Finding the probability distribution which achieves the maximum in \eqref{eq:SECC_Capacity} is not straightforward, in general. If $U_{\mathcal{A}}$ denotes the uniform distribution of $X_1^L$ over $\mathcal{A}$, then the maximum information rate achievable with $U_{\mathcal{A}}$, denoted $C_{U_{\mathcal{A}}}^L(B)$, acts as a lower bound for $C_{SECC}^{L}(B)$. Since a CSCC can be viewed as a SECC where the input vectors have the same composition, it follows that $C_{CSCC}^L(B)$ is also a lower bound for $C_{SECC}^{L}(B)$. Thus we have
\begin{equation}
C_{SECC}^{L}(B) \ge \max\{ C_{CSCC}^L(B) , C_{U_{\mathcal{A}}}^L(B) \} .
\label{eq:SECC_LB}
\end{equation}

The following proposition is useful in reducing the computational complexity of $C_{U_{\mathcal{A}}}^L(B)$.
\begin{proposition}
\label{prop:SECC_PairwiseProb}
For a random input vector $X_1^L$ uniformly distributed over $\mathcal{A}$ with corresponding output vector $Y_1^L$, the pairwise joint probability, for $1 \le i \le L$, satisfies
\begin{equation}
P_{XY}(X_i = x , Y_i = y) = \sum_{P \in \Gamma_{B}^{L}} \frac{|\mathcal{T}_{P}^{L}|}{|\mathcal{A}|} P(x) W(y|x)
\label{eq:SECC_PairwiseProb}
\end{equation}
\end{proposition}
\begin{IEEEproof}
When $X_1^L$ is uniformly distributed over $\mathcal{A}$, 
\begin{equation}
\mathrm{Pr}(X_1^L \in \mathcal{T}_{P}^{L}) = \frac{|\mathcal{T}_{P}^{L}|}{|\mathcal{A}|} .
\label{eq:SECC_InputVectorProb}
\end{equation} 
From Prop.~\ref{prop:PairwiseProb} it follows that
\begin{equation}
\mathrm{Pr}(X_i = x , Y_i = y \, | \, X_1^L \in \mathcal{T}_{P}^{L}) = P(x) W(y|x) .
\label{eq:SECC_InputVectorConditionalProb}
\end{equation}
Finally \eqref{eq:SECC_PairwiseProb} follows from \eqref{eq:SECC_InputVectorProb} and \eqref{eq:SECC_InputVectorConditionalProb} since $P_{XY}(X_i = x , Y_i = y)$ is equal to
\begin{equation}
\sum_{P \in \Gamma_{B}^{L}} \mathrm{Pr}(X_1^L \in \mathcal{T}_{P}^{L}) \ \mathrm{Pr}(X_i = x , Y_i = y \, | \, X_1^L \in \mathcal{T}_{P}^{L}) .
\end{equation}
\end{IEEEproof}

Another useful observation with SECC is that if $y_1^L$ and $\tilde{y}_1^L$ are two output vectors having the same composition, then the columns of the induced vector channel transition matrix corresponding to $y_1^L$ and $\tilde{y}_1^L$ are permutations of each other. This follows from arguments similar to those presented in Appendix~\ref{app:UniformDistInputVectors} for CSCC. Thus, if $X_1^L$ is distributed uniformly over $\mathcal{A}$, then for  $y_1^L$ and $\tilde{y}_1^L$ having the same composition, we have
\begin{equation}
P_{Y_1^L}(\tilde{y}_1^L) = P_{Y_1^L}(y_1^L) = \frac{1}{|\mathcal{A}|} \sum_{x_1^L \in \mathcal{A}} W^L(y_1^L |x_1^L) .
\label{eq:SECC_Unif_OutputProbDist}
\end{equation}
The next proposition gives a computationally efficient expression for $C_{U_{\mathcal{A}}}^L(B)$.
\begin{proposition}
$C_{U_{\mathcal{A}}}^L(B)$ can be expressed as
\begin{equation}
\frac{1}{L} \sum_{Q \in \mathcal{Q}_L} |\mathcal{T}_{Q}^{L}| \, P_{Y_1^L}(y_1^L) \log \frac{1}{P_{Y_1^L}(y_1^L)} \ - \ H(Y|X) ,
\label{eq:SECC_UniformDistCapacity}
\end{equation}
where $\mathcal{Q}_L$ is the set of all compositions for output vectors of length $L$,
only one representative output vector $y_1^L$ is chosen from every type class $\mathcal{T}_{Q}^{L}$, $P_{Y_1^L}(y_1^L)$ is given by \eqref{eq:SECC_Unif_OutputProbDist}, and
$H(Y|X)$ is evaluated using the joint pairwise probability distribution given by \eqref{eq:SECC_PairwiseProb}.
\end{proposition}
\begin{IEEEproof}
For a DMC, we have 
\begin{equation}
C_{U_{\mathcal{A}}}^L(B) = \frac{1}{L} \left( H(Y_1^L) - \sum_{i=1}^L H(Y_i | X_i) \right) ,
\label{eq:SECC_UniformDistCapacity2}
\end{equation}
where the probability of $y_1^L \in \mathcal{Y}^L$ is given by \eqref{eq:SECC_Unif_OutputProbDist}. Thus, \eqref{eq:SECC_UniformDistCapacity} follows from \eqref{eq:SECC_UniformDistCapacity2}, \eqref{eq:SECC_PairwiseProb} and the observation that output vectors with the same composition have equal probability when input subblocks are uniformly distributed over $\mathcal{A}$.
\end{IEEEproof}

As discussed earlier, the energy requirement per subblock is stricter than the average energy requirement per codeword. Hence, the capacity using codes with subblock-constraint \eqref{eq:SubblockEnergyConstraint} is less than the capacity using codes with codeword constraint \eqref{eq:CodewordEnergyConstraint}. Since CCCs achieve capacity with codeword constraint \cite{CsiszarBook_IIed}, we have
\begin{equation}
C_{SECC}^{L}(B) \le C_{CCC}(B) .
\label{eq:SECC_UB}
\end{equation}
From \eqref{eq:SECC_LB} and \eqref{eq:SECC_UB} it follows that $C_{CSCC}^L(B) \le C_{SECC}^L(B) \le C_{CCC}(B)$. Further, using \eqref{eq:C_CSCC_L_infty} it follows that SECC capacity tends to CCC capacity as $L \to \infty$. We will compare these capacities for different cases in the numerical results section.

\section{Real-time Information Transfer}
\label{Sec:RealTimeInfo}
So far, we could ensure real-time energy transfer to the receiver by placing constraints on the subblock-composition. For information transfer, although joint decoding of all the subblocks within a codeword is preferred for reducing the probability of error, it also causes delay in information arrival. 

For enabling real-time information transfer, the receiver may decode each subblock independently, and thus avoid waiting for arrival of future subblocks. Here, since the subblock decoding proceeds the instant that subblock has been completely received, the information transfer delay is only due to subblock transfer time and the corresponding decoding delay.

When each subblock within the transmitted sequence is decoded independent of other subblocks, then each subblock may itself be viewed as a codeword. We will refer to the independent decoding of subblocks as \emph{local subblock decoding} (LSD). We remark that this subblock based decoding is distinct from decoding for locally decodable codes that allows any bit of the message to be decoded with high probability by only querying a small number of received bits \cite{Yekhanin11}.

\subsection{Local Subblock Decoding}
\label{Sec:LSD}
In case of local subblock decoding, each subblock may be treated as an independent codeword since every subblock is decoded independently. We are interested in estimating achievable rates with bounded error probability when local subblock decoding is employed. We now provide a short review of an existing result on achievable rates for constant composition finite blocklength codes. This result will then be used (in Sec.~\ref{Sec:NumericalResults}) to compare rates between local (independent) subblock decoding and joint subblock decoding.

Let $M^*(n,\epsilon)$ denote the maximum size of length-$n$ constant composition code for a DMC with average error probability no larger than $\epsilon$. 
When the composition of codewords is equal to an input probability distribution which maximizes the mutual information and the channel satisfies some regularity conditions, then \cite{PolyanskiyThesis,Moulin12,Tomamichel13_IT}
\begin{equation}
\log M^*(n,\epsilon) = n C - \sqrt{nV}Q^{-1}(\epsilon) + \frac{1}{2}\log n + O(1)
\end{equation}
where $C$ is the channel capacity, $V$ is the {\emph{information variance}}, and $Q$ is the Gaussian $Q$-function \cite{Moulin12}. We remark that $V$ is also termed {\emph{channel dispersion}} in literature \cite{PPV10}. Early results on finite blocklength capacity for memoryless symmetric channels are due to Weiss~\cite{Weiss60}, which were generalized for the DMC and strengthened by Strassen~\cite{Strassen62}.

When each codeword has equal number of ones and zeros, the achievable rate in bits per channel use for BSC with crossover probability $p$ using CCC is approximated as~\cite{PolyanskiyThesis}:
\begin{equation}
\frac{\log_2 M^*(n,\epsilon)}{n} \approx C - \sqrt{\frac{p(1-p)}{n}} \log_2 \frac{1-p}{p} \, Q^{-1}(\epsilon) + \frac{1}{2n}\log_2 n , 
\label{eq:ApproxRateBSC}
\end{equation}
with $C = 1 + p\log_2 p + (1-p)\log_2 (1-p)$.

\section{Numerical Results and Discussion}
\label{Sec:NumericalResults}

In this section, we provide examples highlighting the tradeoff between delivery of sufficient energy to the receiver and achieving high information transfer rates. These results are used to draw meaningful insights into choice of subblock length and subblock composition as a function of required energy per symbol at the receiver.

\begin{figure}
\centering
\includegraphics[width=\myfigwidth\textwidth]{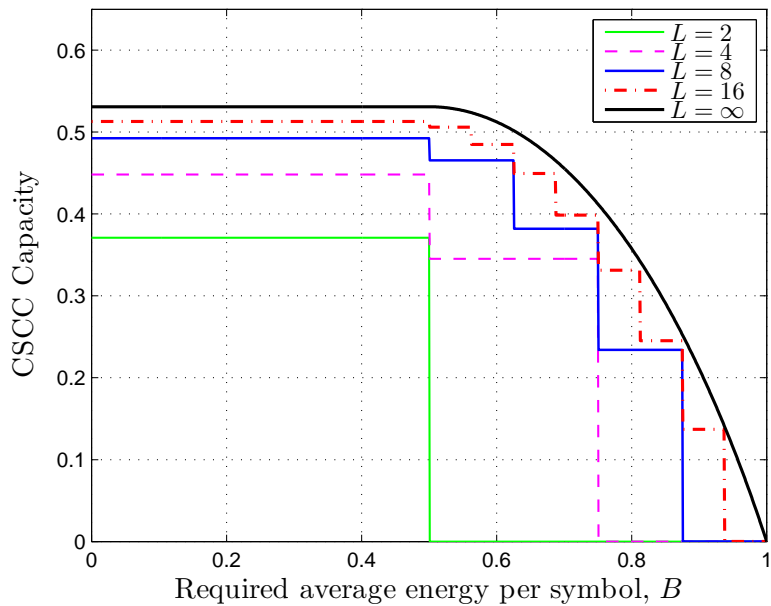}
\caption{Plot of $C_{CSCC}^L(B)$ versus $B$ for BSC with crossover probability $p_0 = 0.1, \, b(0)=0,  \, b(1)=1$.}
\label{Fig:CSCC_CapacityEnergyFn_v4}
\end{figure}

Fig.~\ref{Fig:CSCC_CapacityEnergyFn_v4} plots $C_{CSCC}^L(B)$ as a function of $B$ for different values of $L$ for a BSC with crossover probability $p_0 = 0.1$. The $b$-values are assumed to be $b(0)=0$ and $b(1)=1$. These $b$-values reflect the case of on-off keying where bit-1 (bit-0) is represented by the presence (absence) of a carrier signal. Fig.~\ref{Fig:CSCC_CapacityEnergyFn_v4} shows that, in general, the value of information rate given by $C_{CSCC}^L(B)$ increases with an increase in the subblock length $L$, for a given $B$. This is because an increase in $L$ leads to greater choice for input symbols within a subblock. Note that the smaller the value of $L$, the greater the uniformity in energy distribution within a codeword. The reduction in capacity due to choice of smaller $L$ is the price we pay for providing smoother energy content.

The plot for $L=\infty$ is evaluated using \eqref{eq:ConstrainedCapacity}; this follows from \eqref{eq:CSCC_Capacity}, \eqref{eq:Diff_CSCC_CCC}, \eqref{eq:CCC_B_Capacity}, and the fact that
$\lim_{L \to \infty} r(L,P) = 0$. Thus the curve corresponding to $L = \infty$ is same as the $C_{CCC}(B)$ curve. This curve is a non-increasing concave function of $B$ for $0 \le B \le b_{\max}$. This claim can be proved using the approach in \cite{Lav08}. It is non-increasing since the feasibility set $\Gamma_{B}^{L}$ will only become smaller on increasing $B$. The concavity of $C_{CCC}(B)$ follows from the concavity of $I(X;Y)$ as a function of probability distribution of $X$ and the fact that for $0 < \alpha < 1$, the conditions $\mathbb{E}_{P_1}[b(X)] \ge B_1$ and $\mathbb{E}_{P_2}[b(X)] \ge B_2$ imply that
\begin{equation}
\mathbb{E}_{\alpha P_1 + (1 - \alpha) P_2}[b(X)] \ge \alpha B_1 + (1 - \alpha) B_2 .
\end{equation}
The non-increasing concave nature of the capacity-power function was used in \cite{Lav12} to show the suboptimality of a time-sharing approach to energy and information transfer.

\begin{figure}
\centering
\includegraphics[width=\myfigwidth\textwidth]{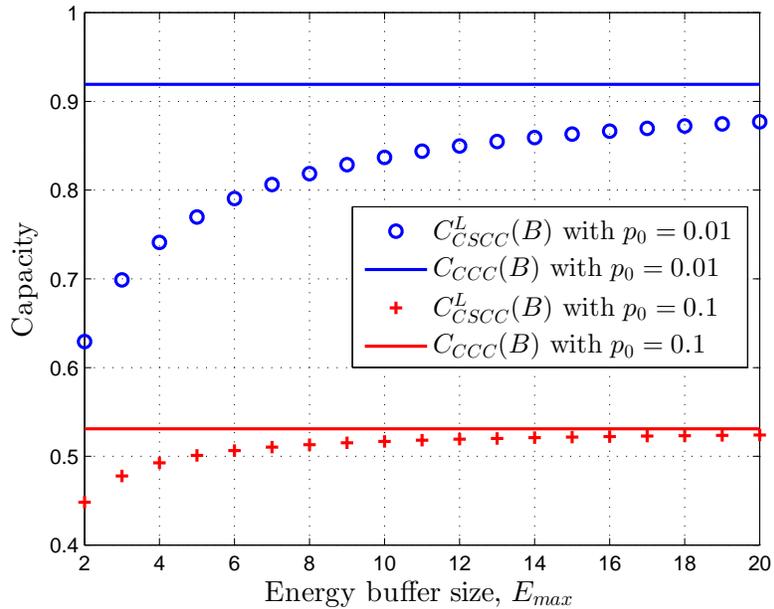}
\caption{Plot of CSCC capacity versus receiver energy buffer size, $E_{max}$, with $B=0.5, b(0)=0, b(1)=1$ for BSC with crossover probability $p_0 = \{0.01, 0.1\}$.}
\label{Fig:CSCC_CapacityVsEmax}
\end{figure}

The CSCC capacity is plotted in Fig.~\ref{Fig:CSCC_CapacityVsEmax} for a BSC as a function of the receiver energy buffer size, $E_{max}$, with $B=0.5$. The subblock length $L$ is chosen as a function of $E_{max}$ to satisfy \eqref{eq:BoundOnL}. Since $L$ increases with increasing values of $E_{max}$, the CSCC capacity is an increasing function of $E_{max}$. For $p_0 = 0.1$, the CSCC capacity is limited by the relatively high value of the crossover probability, rather than the subblock length, with capacity remaining almost constant as $E_{max}$ is increased beyond 10. On the other hand, for $p_0 = 0.01$, the CSCC capacity is limited by the subblock length (since `noise' is weak). From \eqref{eq:BoundOnL} we observe that the subblock length tends to infinity as $E_{max}$ tends to infinity, and hence the CSCC capacity corresponding to $E_{max} \to \infty$ is equal to $C_{CCC}(B)$.

\begin{figure}
\centering
\includegraphics[width=\myfigwidth\textwidth]{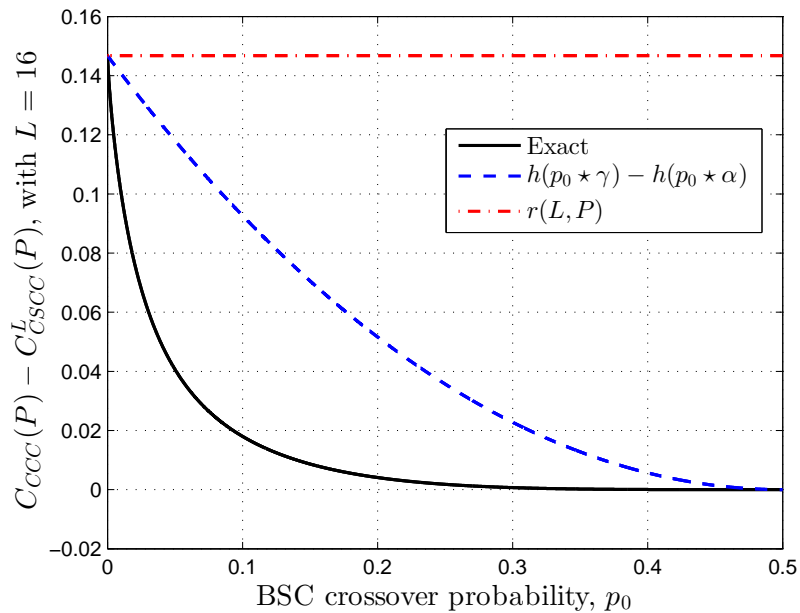}
\caption{Plot of $C_{CCC}(P) - C_{CSCC}^{L}(P)$ as a function of BSC crossover probability $p_0$ for $L=16$ and $\Pr(0)=\Pr(1)=0.5$.}
\label{Fig:RatePenalty_v1}
\end{figure}

Fig.~\ref{Fig:RatePenalty_v1} plots the rate penalty incurred by using CSCC instead of CCC, for a BSC with crossover probability $p_0$, $L=16$, and $\Pr(0)=\Pr(1)=0.5$. As discussed in Sec.~\ref{subsec:RatePenalty}, the upper bound on the rate penalty given by $r(L,P)$ is shown to be close to the exact value when $p_0 \approx 0$. Note that $r(L,P)$ is independent of the underlying channel. A tighter bound on the rate penalty given by $h(p_0 \star \gamma) - h(p_0 \star \alpha)$ is also plotted (see Theorem~\ref{th:RateLossBound}). These bounds are useful in estimating the rate penalty for large values of $L$ when the computational complexity of $C_{CSCC}^{L}(P)$ becomes high. The bounds on rate penalty may also be used to bound the exact value of $C_{CSCC}^{L}(P)$ for large $L$.

\begin{figure}
\centering
\includegraphics[width=\myfigwidth\textwidth]{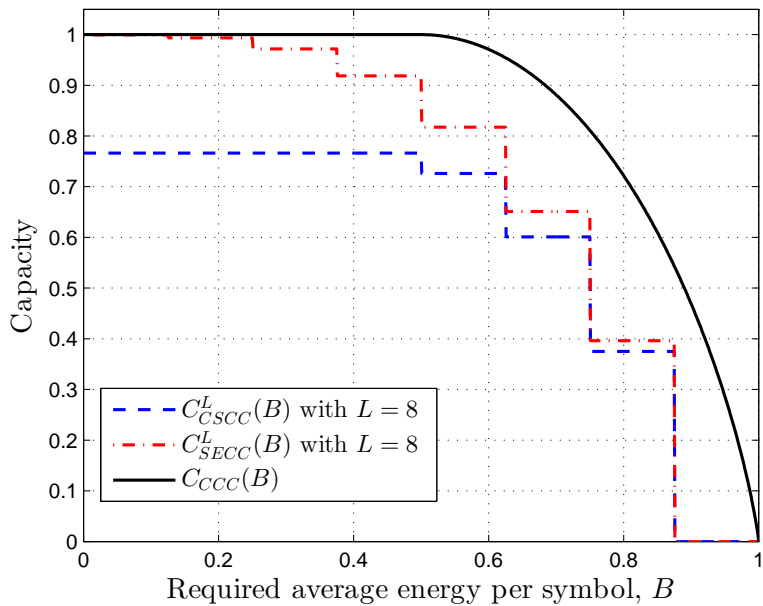}
\caption{Comparison of capacity of different schemes for a noiseless binary channel with $b(0)=0, b(1)=1$.}
\label{Fig:SECC_CapacityEnergyFn}
\end{figure}

Fig.~\ref{Fig:SECC_CapacityEnergyFn} compares the capacity of CSCC and SECC for a noiseless binary channel with $b(0)=0, b(1)=1$ and subblock length $L=8$. Note that the capacity curve for CCC may be viewed as the CSCC capacity curve corresponding to $L=\infty$. Fig.~\ref{Fig:SECC_CapacityEnergyFn} highlights the potential of improving the CSCC capacity by using SECCs and allowing different subblocks to have different compositions while still meeting the subblock energy constraint \eqref{eq:SubblockEnergyConstraint}. With SECCs, the capacity for a noiseless channel is achieved by a uniform distribution of input vectors and can thus be efficiently computed using \eqref{eq:SECC_UniformDistCapacity}.

\begin{figure}
\centering
\includegraphics[width=\myfigwidth\textwidth]{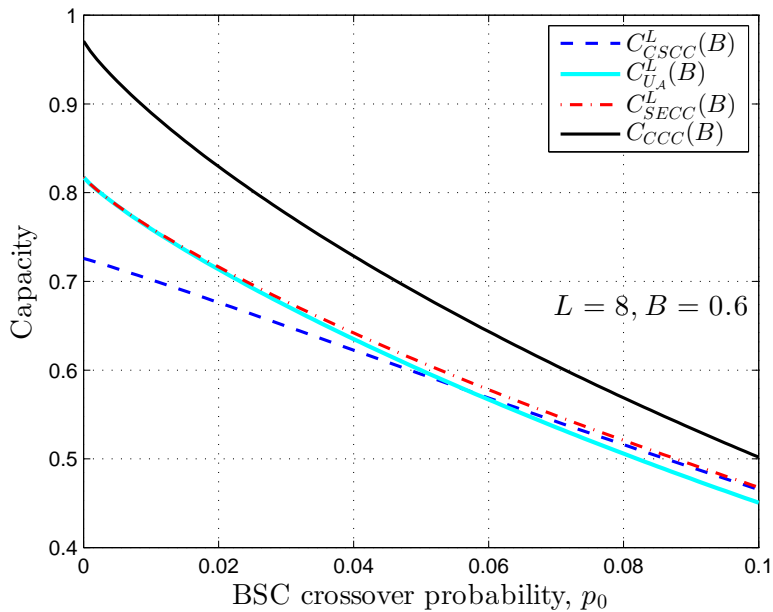}
\caption{Comparison of capacity of different schemes for $L=8, B=0.6$, as a function of BSC crossover probability $p_0$ and $b(0)=0, b(1)=1$.}
\label{Fig:SECC_CapacityVs_p0}
\end{figure}

Fig.~\ref{Fig:SECC_CapacityVs_p0} compares capacity of different schemes for $L=8$ and $B=0.6$, as a function of BSC crossover probability $p_0$. It shows that for $p_0 < 0.05$, the capacity with uniform distribution over the set of length $L$ vectors which satisfy the subblock energy constraint \eqref{eq:SubblockEnergyConstraint}, is higher compared to CSCC capacity. However, $C_{U_{\mathcal{A}}}^L(B) < C_{CSCC}^L(B)$ for relative higher values of $p_0$. This observation emphasizes the fact that merely adding more types is not sufficient to increase capacity compared to CSCC; we need to choose an appropriate distribution over the enlarged alphabet as well. In Fig.~\ref{Fig:SECC_CapacityVs_p0}, we used the Blahut-Arimoto algorithm \cite{Arimoto72,Blahut72} to compute the exact SECC capacity, $C_{SECC}^L(B)$.

\begin{figure}
\centering
\includegraphics[width=\myfigwidth\textwidth]{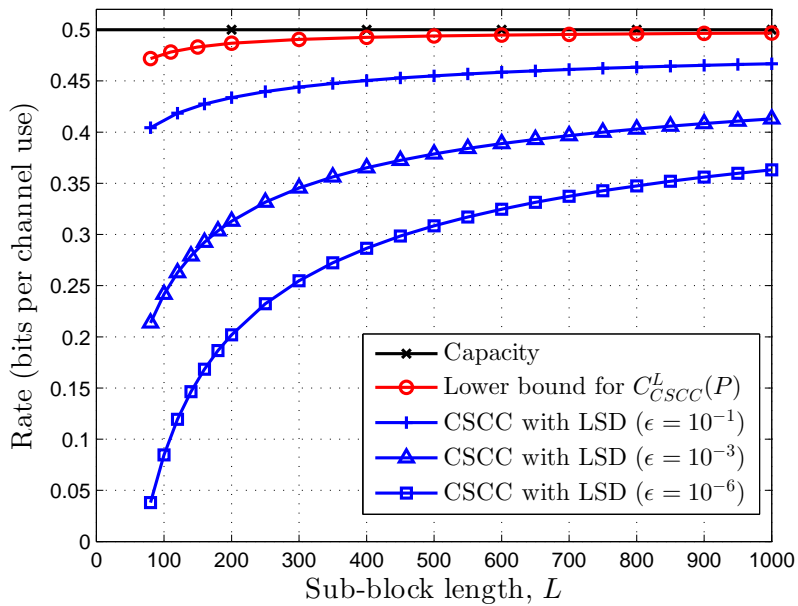}
\caption{Rates for a BSC with crossover probability $p=0.11$.}
\label{Fig:BSC_p_0.11}
\end{figure}

Fig.~\ref{Fig:BSC_p_0.11} compares achievable rates using local subblock decoding (LSD) with rates using joint subblock decoding for a BSC with crossover probability $p_0=0.11$ when each subblock has equal number of zeros and ones (that is, $P(0)=P(1)=0.5$). In case of CSCC with LSD, each subblock may itself be viewed as a codeword and so the achievable rate is approximated by \eqref{eq:ApproxRateBSC} with $n = L$. The achievable rates with LSD are obtained using \eqref{eq:ApproxRateBSC} and seen to fall significantly as the desired probability of error, $\epsilon$, tends to zero. The red curve plots lower bound on $C_{CSCC}^L(P)$ obtained using \eqref{eq:RateLossBound_BSC}. Note that $C_{CSCC}^L(P)$ represents the rate with joint subblock decoding for which the probability of error can be brought arbitrarily close to zero by increasing the number of subblocks in a codeword and then jointly decoding the subblocks.

Notice that the rate loss decreases as $\sqrt{1/L}$ with LSD whereas the rate loss with joint decoding decreases as $\log(L) / L$.  Ensuring the ability to use energy in real-time imposes less of a penalty than the ability to use information in real-time.

\section{Reflections}
\label{Sec:Conclusion}

We proposed the use of CSCC codes for providing regular energy content in a patterned energy signal which is used for simultaneous transfer of energy and information. The subblock-composition in CSCC was chosen to maximize the rate of information transfer while ensuring that the fraction of input symbols carrying high energy within every subblock duration are sufficiently large. For characterizing the exact CSCC capacity, we employed a super-letter approach (with each subblock being viewed as a single super-letter in an induced vector-channel) and showed that CSCC capacity computational complexity can be alleviated by exploiting certain symmetry properties. 

The super-letter approach can also be applied to compute CSCC error exponent. However, the size of the super-alphabet grows exponentially with subblock length, $L$, and the cost for computing exact CSCC capacity and error exponent may become prohibitive for large $L$. In this scenario, the CSCC capacity and error exponent can be estimated by using their respective bounds, derived in Sec.~\ref{Sec:CSCC_vs_CCC}, in terms of the capacity and error exponent for constant composition codes. Compared to CCC, the use of CSCCs incurs a rate loss due to the constraint restricting the subblocks to have the same composition. We showed that the CSCC error exponent is related to the CCC error exponent by the same rate loss term. 

We also showed that CSCC capacity can be increased by allowing different subblocks to have different compositions, while still meeting the subblock energy constraint. Although CSCC capacity is shown to be achieved by uniform distribution of super-letters, one may have to resort to numerical techniques (such as the Blahut-Arimoto algorithm) for obtaining a capacity achieving input distribution for the case where different subblocks are permitted to have different compositions.

We provided examples highlighting the tradeoff between delivery of sufficient energy to the receiver and achieving high information transfer rates. It was observed that the ability to use energy in real-time imposes less of penalty than the ability to use information in real-time.

We showed that the subblock length in CSCC can be bounded as a function of the receiver energy storage capacity to avoid energy outage at the receiver. In scenarios where the energy harvested at the receiver upon transmission of an input symbol varies over time, it will be appealing to analyze bounds on subblock length which apply energy arrival statistics to ensure that the energy outage probability is lower than a certain threshold. Future work may also be carried on extending CSCC capacity results to other channel models, such as the AWGN channel where the average transmit power is also constrained.

Other than the application of simultaneous energy and information transfer, CSCCs are also suitable candidates for power line communications due to their ability to provide regular energy content. The CSCC codes may also find application in other diverse fields. For instance, the multiply constant-weight codes (MCWC) proposed in~\cite{Cherif13} for use in low-cost authentication methods are a special case of CSCC with binary input alphabet. Thus, our capacity results for CSCC can also be employed as a performance benchmark for practical MCWC codes~\cite{Chee14_MCWC}.

\section*{Acknowledgment}
The authors thank Vincent~Y.~F.~Tan for his helpful comments.

\appendices
\section{Proof of Theorem~\ref{th:UniformDistInputVectors}}
\label{app:UniformDistInputVectors}
We will prove Theorem~\ref{th:UniformDistInputVectors} by first proving some simple lemmas and employing Gallager's definition of a symmetric channel \cite{GallagerBook68}.

If $\pi$ denotes any permutation on $L$ letters with
\begin{equation}
\pi(x_1^L) = \pi(x_1, x_2, \ldots , x_L) \triangleq (x_{\pi(1)}, x_{\pi(2)}, \ldots , x_{\pi(L)}) ,
\end{equation}
then we have the following lemmas.

\begin{lemma}
\begin{equation}
W^{L} \left(\pi(y_1^L) | \pi(x_1^L) \right) = W^L \left(y_1^L | x_1^L \right)
\end{equation}
\label{lemma:perm_invariant}
\end{lemma}
\begin{IEEEproof}
For a DMC, we have
\begin{align*}
W^L \left(\pi(y_1^L) | \pi(x_1^L) \right) &= \prod_{i=1}^{L} W \left( y_{\pi(i)} | x_{\pi(i)} \right) \\
&= \prod_{i=1}^{L} W(y_{i} | x_{i} ) = W^L \left( y_1^L | x_1^L \right)
\end{align*}
\end{IEEEproof}

\begin{lemma}
The following sets are equal
\begin{equation}
\{ \pi(x_1^L) | x_1^L \in \mathcal{T}_{P}^{L} \} = \mathcal{T}_{P}^{L}
\label{eq:bijection}
\end{equation}
\label{lemma:perm_invariant_II}
\end{lemma}
\begin{IEEEproof}
A permutation preserves the composition of a sequence. Thus, $\pi$ may be viewed as a map $\pi : \mathcal{T}_{P}^{L} \to \mathcal{T}_{P}^{L}$. This map is injective by definition of a permutation. Since the set $\mathcal{T}_{P}^{L}$ is finite, this map is also surjective and hence \eqref{eq:bijection} follows.
\end{IEEEproof}

\begin{lemma}
The following sets are equal
\begin{equation}
\{ W^{L} \left( \pi(y_1^L) | x_1^L \right) : x_1^L \in \mathcal{T}_{P}^{L} \} = \{ W^{L} \left( y_1^L | x_1^L \right) : x_1^L \in \mathcal{T}_{P}^{L} \}
\label{eq:perm_invariant_III}
\end{equation}
\label{lemma:perm_invariant_III}
\end{lemma}
\begin{IEEEproof}
From Lemma~\ref{lemma:perm_invariant} we have $W^L \left( \pi(y_1^L) | x_1^L \right) = W^L \left( y_1^L | \pi^{-1}(x_1^L) \right)$. Now \eqref{eq:perm_invariant_III} follows from Lemma~\ref{lemma:perm_invariant_II}.
\end{IEEEproof}

Let the composition of the output vector $y_1^L \in \mathcal{Y}^L$ be $Q$ and let $\mathcal{T}_{Q}^{L}$ be the set of all output vectors of length $L$ having composition $Q$.
\begin{lemma}
The following sets are equal
\begin{equation}
\{ W^{L} \left( y_1^L | \pi(x_1^L) \right) : y_1^L \in \mathcal{T}_{Q}^{L} \} = \{ W^{L} \left( y_1^L | x_1^L \right) : y_1^L \in \mathcal{T}_{Q}^{L} \}
\end{equation}
\label{lemma:perm_invariant_IV}
\end{lemma}
\begin{IEEEproof}
Similar to Lemma~\ref{lemma:perm_invariant_III}.
\end{IEEEproof} 

We recall Gallager's definition \cite{GallagerBook68} of a \emph{symmetric} DMC.
\begin{definition}
A DMC is \emph{symmetric} if the set of outputs can be partitioned into subsets in such a way that for each subset the matrix of transition probabilities (using inputs as rows and outputs of the subsets as columns) has the property that each row is a permutation of each other row and each column (if more than 1) is a permutation of each other column.
\end{definition}

We will show that when CSCC is employed on a DMC, the induced vector-channel is symmetric. Note that the underlying (scalar) channel can be any arbitrary DMC (not necessarily symmetric).
\begin{lemma}
When CSCC with subblock length $L$ is employed on any DMC, the induced vector-channel (obtained from $L$ uses of the DMC) is symmetric.
\label{lemma:CSCC_symmetric}
\end{lemma}
\begin{IEEEproof}
The lemma will be proved if we can partition the outputs into subsets such that for each subset the matrix of transition probabilities has the property that each row (column) is a permutation of each other row (column).

We now show that if we partition the outputs into subsets such that each subset contains all the outputs of a given composition, then the symmetry conditions will be satisfied.

If $y_1^L \in \mathcal{T}_{Q}^{L}$ and $\tilde{y}_1^L \in \mathcal{T}_{Q}^{L}$ for a given composition $Q$, then since $y_1^L$ and $\tilde{y}_1^L$ have the same composition, we have $\tilde{y}_1^L = \pi(y_1^L)$ for some permutation $\pi$. Let $\mathcal{T}_{P}^{L}$ be the input alphabet for the induced vector channel using CSCC with subblock-composition $P$. Then using Lemma~\ref{lemma:perm_invariant_III}, we note that the columns of the vector-channel transition matrix corresponding to output subset $\mathcal{T}_{Q}^{L}$ are permutations of each other. Similarly, using Lemma~\ref{lemma:perm_invariant_IV} we can prove that the corresponding rows are permutations of each other.
\end{IEEEproof}

\begin{theorem}[{\cite[p.~94]{GallagerBook68}}]
For a symmetric discrete memoryless channel, capacity is achieved by using the inputs with equal probability.
\label{th:Gallager_symmetric}
\end{theorem}

Finally, Theorem~\ref{th:UniformDistInputVectors} follows directly from Lemma~\ref{lemma:CSCC_symmetric} and Theorem~\ref{th:Gallager_symmetric}.  \hfill $\blacksquare$

\section{Proof of Theorem~\ref{th:BoundOnL}}
\label{app:BoundOnL}

When $L$ satisfies \eqref{eq:BoundOnL}, then $E_{max} \ge 2 G$. Since $E(1) \ge G$, the energy level at the start of \emph{every} subblock is at least $G$ (by recursive application of Lemma~\ref{lemma:NoOutage}$(c)$) and sufficiency follows from Lemma~\ref{lemma:NoOutage}$(b)$.

Now let $L_1 = \sum_{x \in \mathcal{X}_{\triangleleft}} L P(x)$, and define
\begin{align}
P_1(x) &= \begin{cases}
\frac{P(x)}{ \sum_{x \in \mathcal{X}_{\triangleleft}} P(x)} , & \mbox{ if } x \in \mathcal{X}_{\triangleleft} \\
0, & \mbox{ if } x \in \mathcal{X}_{\triangleright}
\end{cases} \\
P_2(x) &= \begin{cases}
0, & \mbox{ if } x \in \mathcal{X}_{\triangleleft} \\
\frac{P(x)}{\sum_{x \in \mathcal{X}_{\triangleright}} P(x)} , & \mbox{ if } x \in \mathcal{X}_{\triangleright}
\end{cases} \\
S_1 &= \{x_1^L \, | \, x_1^{L_1} \in \mathcal{T}_{P_1}^{L_1} , \ x_{L_1+1}^L \in \mathcal{T}_{P_2}^{L - L_1} \} \\
S_2 &= \{x_1^L \, | \, x_1^{L-L_1} \in \mathcal{T}_{P_2}^{L - L_1} , \ x_{L-L_1+1}^L \in \mathcal{T}_{P_1}^{L_1} \} .
\end{align}
Clearly $S_1 \subset \mathcal{T}_{P}^L, \ S_2 \subset \mathcal{T}_{P}^L$, where $S_1$ (resp. $S_2$) denotes the set of subblocks of length $L$ with first (resp. last) $L_1$ input symbols belonging to $\mathcal{X}_{\triangleleft}$. Note that $E(1) \ge G$ is necessary to avoid outage because if $E(1) < G$, then outage results when the first subblock in a codeword belongs to $S_1$. To prove that \eqref{eq:BoundOnL} is necessary, we will show that when
\begin{equation}
L > \frac{E_{max}}{\sum_{x \in \mathcal{X}_{\triangleleft}} 2 P(x) \left(B - b(x)\right)} ,
\label{eq:IncorrectBoundOnL}
\end{equation}
then CSCC codewords exist which will result in energy outage at the receiver. Here we have
\begin{equation}
G = \sum_{x \in \mathcal{X}_{\triangleleft}} L P(x) \left(B - b(x)\right) > \frac{E_{max}}{2} \ .
\label{eq:EnergyIncrement}
\end{equation}
Let the first subblock in a given codeword belong to $S_2$. Since the last $L_1$ symbols (within the first subblock) belong to $\mathcal{X}_{\triangleleft}$, we have $E(L+1) = |E(L - L_1+1) - G|^+$.
If there is no outage during the reception of the first subblock,
\begin{equation}
E(L+1) = E(L - L_1+1) - G \le E_{max} - G < E_{max}/2 ,
\label{eq:EnergyAfterSubblock1}
\end{equation}
where the last inequality follows from \eqref{eq:EnergyIncrement}. Now let the second subblock belong to $S_1$. There is no energy outage during the reception of first $L_1$ symbols within the second subblock if and only if $E(L+1)\ge G$. However, from \eqref{eq:EnergyAfterSubblock1} and \eqref{eq:EnergyIncrement} it follows that $E(L+1) < E_{max}/2 < G$, and hence outage cannot be avoided in the second subblock. In general, outage results if $L$ satisfies \eqref{eq:IncorrectBoundOnL}, and any two adjacent subblocks in a codeword belongs to $S_2$ and $S_1$, respectively. \hfill $\blacksquare$

\section{Proof of Theorem~\ref{th:RateLossBound}}
\label{app:RateLossBound}
The strict inequality $0 < C_{CCC}(P) - C_{CSCC}^{L}(P)$ follows for BSC with crossover probability $0 < p_0 < 0.5$ because
\begin{align}
C_{CSCC}^{L}(P) &= \frac{1}{L} \left[ H(Y_1^L) - H(Y_1^L | X_1^L)\right] \\
&= \frac{1}{L} \left[ \sum_{i=1}^{L} H(Y_i | Y_1^{i-1}) - \sum_{i=1}^{L} H(Y_i | X_i)\right] \\
&\overset{(a)}{<} \frac{1}{L} \left[ \sum_{i=1}^{L} H(Y_i) - \sum_{i=1}^{L} H(Y_i | X_i)\right] \\
&= C_{CCC}(P) ,
\end{align}
where $Y_1^{i-1} = Y_1\ldots Y_{i-1}$, the strict inequality $(a)$ follows since $Y_i$ is related to $Y_{1}^{i-1}$ via $X_{1}^{i-1}$ and $X_i$. The last equality above follows from Prop.~\ref{prop:PairwiseProb} and \eqref{eq:CCC_Capacity}.

For subblock-composition $P$ with  $0 < \gamma = \min(P(0),P(1)) \le 0.5$, the output entropy on a BSC is $H(Y) = h(p_0 \star \gamma)$ and hence
\begin{equation}
C_{CCC}(P) = h(p_0 \star \gamma) - h(p_0) .
\label{eq:CCC_CapacityBSC}
\end{equation} 
For CSCC, from \eqref{eq:TypeClassSize} and definition of $\alpha$, it follows that
\begin{align}
\frac{1}{L} H(X_1^L) &= H(P) - r(L,P) \\
&= h(\gamma) - r(L,P) = h(\alpha) .
\label{eq:InputVectorEntropy}
\end{align}
Now using \eqref{eq:InputVectorEntropy} and applying Mrs. Gerber's Lemma \cite{Wyner73}, 
\begin{equation}
\frac{1}{L} H(Y_1^L) \ge h(p_0 \star \alpha) ,
\label{eq:OutputVectorEntropy}
\end{equation}
and hence
\begin{align}
C_{CSCC}^{L}(P) &= \frac{1}{L} \left[ H(Y_1^L) - \sum_{i=1}^{L} H(Y_i | X_i)\right] \\
&\ge h(p_0 \star \alpha) - h(p_0) .
\label{eq:CSCC_CapacityBSC}
\end{align}
Using \eqref{eq:CCC_CapacityBSC} and \eqref{eq:CSCC_CapacityBSC} we have
\begin{equation}
 C_{CCC}(P) - C_{CSCC}^{L}(P) \le h(p_0 \star \gamma) - h(p_0 \star \alpha)
\end{equation}

We only have to show that $h(p_0 \star \gamma) - h(p_0 \star \alpha) < r(L,P)$ for completing the proof. Towards this we first observe that when $0 < x \le 0.5$ and $0 < p_0 < 0.5$, then $p_0 \star x \ge x$. Next we note that the derivative of $h(x)$ satisfies
\begin{equation}
h'(x) = \log \frac{1-x}{x} ,
\end{equation}
and hence $h'(x)$ is a monotonically decreasing function of $x$ for $0 < x \le 0.5$. 

Since $h(\alpha) = h(\gamma) - r(L,P)$, we have 
\begin{align}
h(p_0 \star \gamma) - h(p_0 \star \alpha) &< r(L,P) \iff \nonumber \\
&h(p_0 \star \gamma) - h(\gamma) < h(p_0 \star \alpha) - h(\alpha) .
\label{eq:RateLossBSC}
\end{align}
If we define $f(x) = h(p_0 \star x) - h(x)$ for $0 \le x \le 0.5$, then we have
\begin{equation}
f'(x) = (1-2p_0) h'(p_0 \star x) - h'(x) .
\end{equation}
Hence $f'(x) < 0$ for $0 < x \le 0.5$ since $h'(x)$ is monotonically decreasing in $x$ and $p_0 \star x \ge x$. This in turn implies that $f(x)$ is a strictly monotonically decreasing function of $x$. It follows that $f(\gamma) < f(\alpha)$ (since $\alpha < \gamma$) and \eqref{eq:RateLossBSC} is satisfied. \hfill $\blacksquare$

\section{Proof of Theorem~\ref{th:RateLossBound_BEC}}
\label{app:RateLossBound_BEC}
For a BEC with erasure probability $\epsilon$, and $\gamma = P(0)$,
\begin{equation}
C_{CCC}(P) = (1-\epsilon) h(\gamma) .
\label{eq:CCC_CapacityBEC}
\end{equation}
If $\alpha$ is chosen such that $h(\alpha) = h(\gamma) - r(L,P)$, then from \eqref{eq:TypeClassSize} it follows that $H(X_1^L)/L = h(\alpha)$. Now applying an extension of MGL for binary input symmetric channels \cite{Chayat89}, we get $H(Y_1^L)/L \ge (1-\epsilon) h(\alpha) + h(\epsilon)$. Thus,
\begin{equation}
C_{CSCC}^{L}(P) = \frac{1}{L} \left[ H(Y_1^L) - \sum_{i=1}^{L} H(Y_i | X_i)\right] \ge (1-\epsilon) h(\alpha),
\label{eq:CSCC_CapacityBEC}
\end{equation}
and \eqref{eq:RateLossBound_BEC} follows from \eqref{eq:CCC_CapacityBEC}, \eqref{eq:CSCC_CapacityBEC}, and definition of $\alpha$.

\section{Proof of Theorem~\ref{th:RateLossBound_Z}}
\label{app:RateLossBound_Z}
For a $Z$-channel with $\gamma = \Pr(X=1), \ p_0 = \Pr(1\rightarrow 0)$,
\begin{equation}
C_{CCC}(P) = h\left( \gamma(1-p_0) \right) - \gamma h(p_0).
\label{eq:CCC_CapacityZ}
\end{equation}
If $0 \le \alpha \le 0.5$ is chosen such that $h(\alpha) = h(\gamma) - r(L,P)$, then from \eqref{eq:TypeClassSize} it follows that $H(X_1^L)/L = h(\alpha)$. Now applying the extension of MGL for memoryless asymmetric binary-input, binary-output channels~\cite{Ahlswede77}, we get $H(Y_1^L)/L \ge h\left(\alpha(1-p_0) \right)$. Thus,
\begin{equation}
C_{CSCC}^{L}(P) = h\left( \alpha(1-p_0) \right) - \gamma h(p_0),
\label{eq:CSCC_CapacityZ}
\end{equation}
and \eqref{eq:RateLossBound_Z} follows from \eqref{eq:CCC_CapacityZ} and \eqref{eq:CSCC_CapacityZ}.

\section{Proof of Lemma~\ref{lemma:StructureOfV}}
\label{app:StructureOfV}
We first note that the functions $D(V||W|P)$ and $I(P,V)$ are convex functions of $V$, while the constraint $\sum_{y \in \mathcal{Y}} V(y|x) = 1$ for $x \in \mathcal{X}$ is linear in $V$. Thus the problem of minimization of $D(V||W|P)$ over $V$ subject to $I(P,V) \le R$ and  $\sum_{y \in \mathcal{Y}} V(y|x) = 1$ is a convex optimization problem and can be solved by the method of Lagrange multipliers \cite{BoydBook}.

Secondly, note that the functions $D(V||W|P)$ and $I(P,V)$ depend on $V$ only through those $V(y|x)$ for which $P(x) > 0$. Thus we assume, without loss of generality, that $P(x) > 0, \forall x \in \mathcal{X}$. 

Now consider the Lagrangian $\xi(V)$ given by
\begin{align}
\xi(V) = D(V||W|P) &+ \lambda (I(P,V) - R) \nonumber \\
&+ \sum_x \nu_x \left(\sum_y V(y|x) - 1\right)
\end{align}
where $\lambda \ge 0$. On setting the partial derivative of $\xi(V)$ with respect to $V(y|x)$ equal to zero, we get
\begin{align}
0 = P(x) &\log \frac{V(y|x)}{W(y|x)} + P(x) \log e \nonumber \\
&+ \lambda P(x) \log \frac{V(y|x)}{PV(y)} + \nu_x .
\end{align}
On substituting $s = \lambda/(1+\lambda)$, the above equation can be equivalently be expressed as
\begin{equation}
V(y|x) = W(y|x)^{1-s} PV(y)^s \exp( - \frac{(1-s)(P(x) \log e + \nu_x)}{P(x)})
\label{eq:OptV_v1}
\end{equation}
Since $\sum_y V(y|x) = 1$, using \eqref{eq:OptV_v1} we get
\begin{equation}
\exp\left(\frac{(1-s)(P(x) \log e + \nu_x)}{P(x)}\right) = \sum_y W(y|x)^{1-s} PV(y)^s ,
\label{eq:OptV_v2}
\end{equation}
and finally using \eqref{eq:OptV_v1}, \eqref{eq:OptV_v2}, we have
\begin{equation}
V(y|x) = \frac{W(y|x)^{1-s} PV(y)^s}{\sum_y W(y|x)^{1-s} PV(y)^s} .
\label{eq:OptV_v3}
\end{equation}
Note that since $s = \lambda/(1+\lambda)$, the value of $s$ ranges from $0$ to $1$ as $\lambda$ varies from $0$ to $\infty$. By the complementary slackness property \cite{BoydBook}, we have 
\begin{equation}
\lambda (I(P,V) - R) = 0 .
\label{eq:KKT_CS}
\end{equation}
Since $\lambda = 0$ implies $s = 0$ and hence $V=W$ (using \eqref{eq:OptV_v3}), it follows that $I(P,W) = I(P,V) \le R$. This contradicts the assumption in Lemma~\ref{lemma:StructureOfV} that $R < I(P,W)$ and hence $\lambda$ is strictly greater than zero. The proof is complete by noting that conditions $\lambda > 0$ and \eqref{eq:KKT_CS} imply $I(P,V) = R$. \hfill $\blacksquare$

\section{Proof of Theorem~\ref{th:CSCC_ErrorProb}}
\label{app:CSCC_ErrorProb}
The $M$ messages to be transmitted are assumed equiprobable. All input sequences of length $n$ with constant subblock-composition $P$ are assigned equal probabilities and the $i$th  message is mapped to a randomly selected input sequence for $1 \le i \le M$. The decoder knows the mapping used by the encoder and uses maximum likelihood (ML) decoding.

The proof uses Fano's approach \cite{FanoBook61} to upper bound the probability of error by employing \emph{tilted distribution} which is summarized next. 

Let $A$ be a discrete ensemble consisting of points $a_1, \ldots , a_m$ with probability distribution $\tilde{P}(a)$. If $\phi$ denotes a random variable associated with this ensemble, and $\gamma(s) := \log \sum_A \exp\left(s \phi(a)\right) \tilde{P}(a)$, then a family of \emph{tilted distributions} are: 
\begin{equation}
Q(a) := \exp\left(s \phi(a) - \gamma(s)\right) \tilde{P}(a) \mbox{.}
\end{equation}
Note that for a fixed $s$, the derivative $\gamma'(s)$ (resp. $\gamma''(s)$) denotes the mean (resp. variance) of the random variable $\phi$ with respect to the tilted probability distribution $Q$. For $s=0$, we have $Q=P$, and thus $\gamma'(0)$ (resp. $\gamma''(0)$) denotes the true mean (resp. variance) of $\phi$.

Let $n_i, i = 1,\ldots, K$ be positive integers and $n = \sum_{i=1}^K n_i$. Define the subsets
$S_1 = \{1, \ldots ,n_1\}$ and $S_k = \{ l \in \mathbb{N} \, | \, \sum_{i=1}^{k-1} n_i < l \le \sum_{i=1}^{k} n_i \}$ for $2 \le k \le K$. Let $\alpha^n := \alpha_1 \cdots \alpha_n$ denote a sequence of $n$ independent events. For any $i \in S_k, k = 1,\ldots, K$, let $\phi_k(\alpha_i)$ be the random variable associated with the event $\alpha_i$ and $\tilde{P}_k(a)$ be the probability $\Pr[\alpha_i = a]$. For $k = 1,\ldots,K$, define
\begin{align}
\gamma_k(s) &:= \log \sum_A \exp\left(s \phi_k(a)\right) \tilde{P}_k(a) \\
\gamma(s) &:= \sum_{k=1}^K \frac{n_k}{n} \gamma_k(s) \\
Q_k(a) &:= \exp\left(s \phi_k(a) - \gamma_k(s)\right) \tilde{P}_k(a)
\end{align}

Define the sum of random variables,
\begin{equation}
\Phi(\alpha^n) := \sum_{k=1}^K \, \sum_{i \in S_k} \phi_k(\alpha_i).
\end{equation}
whose tail probability is given by the following lemma.
\begin{lemma}(\cite[p.~265]{FanoBook61}) Assume $\gamma(s)$ and its first and second derivative are finite in the interval $s_1 < s < s_2$ including $s=0$. If $t$ is a real number  with $\gamma'(s_1) < t < \gamma'(s_2)$, then the tail probabilities of $\Phi(\alpha_n)$ satisfy the following inequalities:
\begin{align}
\Pr[\Phi(\alpha^n) \le n t] &\le \exp\left(- n \beta\right), \ \ \gamma'(s_1) < t < \gamma'(0)=\bar{\phi} \\
\Pr[\Phi(\alpha^n) > n t] &\le \exp\left(- n \beta\right), \ \ \gamma'(0)=\bar{\phi} \le t < \gamma'(s_2)
\label{eq:ProbErrorTiltedDist}
\end{align} 
where 
\begin{equation}
\beta = s \gamma'(s) - \gamma(s) = \sum_{k=1}^K \frac{n_k}{n} \,\sum_A Q_k(a) \log \frac{Q_k(a)}{\tilde{P}_k(a)} \ge 0
\end{equation}
with $s$ chosen such that $\gamma'(s) = t$.
\label{lem:TiltedDistribution}
\end{lemma}

Define the \emph{distance} between $x \in \mathcal{X}$ and $y \in \mathcal{Y}$ as 
\begin{equation}
\tilde{D}(x,y) = \log \frac{f(y)}{W(y|x)},
\label{eq:D_def}
\end{equation}
where $f(y)$ is a positive function of $y$ with $\sum_{\mathcal{Y}} f(y) = 1$. Similarly, the distance between two sequences $\mathbf{u}$ and $\mathbf{v}$ is
\begin{equation}
\tilde{D}({\bf{u,v}}) = \sum_{\cal{X,Y}} n(x,y) \tilde{D}(x,y) = \log \frac{F(\mathbf{v})}{W^n(\mathbf{v}|\mathbf{u})}
\label{eq:DistUV}
\end{equation}
where $n(x,y)$ is the number of letter pairs $(x,y)$ in $(\mathbf{u},\mathbf{v})$,
\begin{equation}
F(\mathbf{v}) = \prod_{\cal{Y}} {f(y)}^{n(y)} \mbox{ \ , \ } W^n(\mathbf{v}|\mathbf{u}) = \prod_{\cal{X,Y}} {W(y|x)}^{n(x,y)} \mbox{.}
\label{eq:CapitalF_Def}
\end{equation}
If $\mathbf{u}$ is the input sequence and $\mathbf{v}$ is the corresponding output sequence, then an error may occur with ML decoding when one of the other $M-1$ messages is represented by another input sequence $\mathbf{u'}$ for which $W^n(\mathbf{v}|\mathbf{u'}) \ge W^n(\mathbf{v}|\mathbf{u})$, or equivalently $\tilde{D}(\mathbf{u',v}) \le \tilde{D}(\mathbf{u,v})$. The following lemma gives an upper bound on the probability of error.
\begin{lemma}(\cite[p.~307]{FanoBook61})
Let $D_0$ be an arbitrary constant, $\mathbf{u_0}$ be a particular transmitted codeword, $\mathbf{u}$ be another randomly chosen input sequence, and $\mathbf{v}$ be an output sequence. The average probability of error satisfies the inequality
$P_e < M P_1 + P_2$, where:
\begin{align}
P_1 &= Pr[\tilde{D}(\mathbf{u_0,v}) \le D_0, \tilde{D}(\mathbf{u,v}) \le \tilde{D}(\mathbf{u_0,v})] , \label{eq:BoundP1} \\
P_2 &= Pr[\tilde{D}(\mathbf{u_0,v}) > D_0] \label{eq:BoundP2}. 
\end{align}
For CCC, $P_e$ is independent of $\mathbf{u_0}$.
\label{lem:AvProbUB}
\end{lemma}
Since a CSCC is also a CCC, Lemma~\ref{lem:AvProbUB} will be used to bound the error probability for CSCC, while Lemma~\ref{lem:TiltedDistribution} will be used to compute $P_2$~\eqref{eq:BoundP2}.

We now define some terms and notation which will be used later. We say that sequences $\mathbf{v}$ and $\mathbf{v'}$, each comprising of $m$ subblocks of length $L$, have the \emph{same subblock-composition} if the $i$th  subblock in sequences $\mathbf{v}$ and $\mathbf{v'}$ has the same composition, for $1 \le i \le m$. The composition of the $i$th  subblock of a sequence pair $(\mathbf{u,v})$ is defined as a matrix whose $(j,k)$ entry is equal to $n_i(x_j,y_k)/L$ where $n_i(x_j,y_k)$ is the number of letter pairs $(x_j,y_k)$ in the $i$th  subblock of $(\mathbf{u,v})$. The \emph{subblock-composition of a sequence pair} $(\mathbf{u,v})$ is defined as a length $m$ vector whose $i$th  entry is the composition of the $i$th  subblock of $(\mathbf{u,v})$.

The following lemma compares distances between sequences having the same subblock-composition. This lemma will be used to bound $P_1$~\eqref{eq:BoundP1}.
\begin{lemma}
Let $\mathbf{u}$ and $\mathbf{v}$ be two particular sequences with elements in $\mathcal{X}$ and $\mathcal{Y}$, respectively. Select equiprobably at random a sequence $\mathbf{u'}$ having the same subblock-composition as $\mathbf{u}$, and a sequence $\mathbf{v'}$ having the same subblock-composition as $\mathbf{v}$. Then
\begin{equation}
\Pr[\tilde{D}(\mathbf{u',v}) \le \tilde{D}(\mathbf{u,v})] = \Pr[\tilde{D}(\mathbf{u,v'}) \le \tilde{D}(\mathbf{u,v})]
\label{eq:TildeUEQTildeV}
\end{equation}
\label{lemma:ReplaceUWithV}
\end{lemma} 
\begin{IEEEproof}
Let $n_i(x)$ (resp. $n_i(y)$) denote the number of occurrences of $x$ (resp. $y$) in the $i$th  subblock of $\mathbf{u}$ (resp. $\mathbf{v}$). Let $T_{\mathbf{u,v}}$ be the set of distinct subblock-compositions for sequence pairs $(\mathbf{{u'},{v'}})$ for which $\tilde{D}(\mathbf{{u'},{v'}}) \le \tilde{D}(\mathbf{u,v})$ and $\mathbf{u'}$ (resp. $\mathbf{v'}$) has the same subblock-composition as $\mathbf{u}$ (resp. $\mathbf{v}$).

The number of sequences $\mathbf{{u'}}$, having the same subblock-composition as $\mathbf{{u}}$, for which $\tilde{D}(\mathbf{{u'},v}) \le \tilde{D}(\mathbf{u,v})$ is 
\begin{equation}
\mathlarger{\mathlarger{\sum_{T_{\mathbf{u,v}}}}} \ \prod_{i=1}^{m} \left(\frac{ \displaystyle \prod_{\mathcal{Y}} n_i(y)!}{ \displaystyle \prod_{\mathcal{X,Y}} n_i(x,y)!} \right) 
\label{eq:CSCC_Lemma_Num}
\end{equation}
The total number of sequences $\mathbf{{u'}}$ having the same subblock-composition as $\mathbf{{u}}$ is
\begin{equation}
\prod_{i=1}^{m} \left( \frac{L!}{\displaystyle \prod_{\mathcal{X}} n_i(x)!} \right) = \frac{\left(L!\right)^m}{\displaystyle \prod_{i=1}^{m} \prod_{\mathcal{X}} n_i(x)!}
\label{eq:CSCC_Lemma_Denom}
\end{equation}
Thus, $\Pr[\tilde{D}(\mathbf{u',v}) \le \tilde{D}(\mathbf{u,v})]$ is equal to the ratio of \eqref{eq:CSCC_Lemma_Num} to \eqref{eq:CSCC_Lemma_Denom}.

The number of sequences $\mathbf{{v'}}$, having the same subblock-composition as $\mathbf{{v}}$, for which $\tilde{D}(\mathbf{u,{v'}}) \le \tilde{D}(\mathbf{u,v})$ is 
\begin{equation}
\mathlarger{\mathlarger{\sum_{T_{\mathbf{u,v}}}}} \ \prod_{i=1}^{m} \left(\frac{ \displaystyle \prod_{\mathcal{X}} n_i(x)!}{ \displaystyle \prod_{\mathcal{X,Y}} n_i(x,y)!} \right)
\label{eq:CSCC_Lemma_Num2}
\end{equation}
The total number of sequences $\mathbf{{v'}}$ having the same subblock-composition as $\mathbf{{v}}$ is
\begin{equation}
\prod_{i=1}^{m} \left( \frac{L!}{\displaystyle \prod_{\mathcal{Y}} n_i(y)!} \right) = \frac{\left(L!\right)^m}{\displaystyle \prod_{i=1}^{m} \prod_{\mathcal{Y}} n_i(y)!}
\label{eq:CSCC_Lemma_Denom2}
\end{equation}

Thus, $\Pr[\tilde{D}(\mathbf{u,v'}) \le \tilde{D}(\mathbf{u,v})]$ is the ratio of \eqref{eq:CSCC_Lemma_Num2} to \eqref{eq:CSCC_Lemma_Denom2} which is equal to the ratio of \eqref{eq:CSCC_Lemma_Num} to \eqref{eq:CSCC_Lemma_Denom}.
\end{IEEEproof}

We now proceed with the main steps leading to the proof of Theorem~\ref{th:CSCC_ErrorProb}.

From Lemma~\ref{lem:AvProbUB} it follows that the probability of error (averaged over mappings from messages to codewords) satisfies $P_e < M P_1 + P_2$, where $P_2$ can be bounded using Lemma~\ref{lem:TiltedDistribution} so that $P_2 < \exp(-n \beta)$, with 
\begin{align}
\beta &= s \gamma'(s) - \gamma(s) = D(V||W | P) , \label{eq:Beta_Def} \\
\gamma(s) &=\sum_{\mathcal{X}} P(x) \ \log \sum_{\mathcal{Y}} \exp\left(s \tilde{D}(x,y)\right) W(y|x) ,\label{eq:gammas_val} \\
V(y|x) &= \frac{\exp\left(s \tilde{D}(x,y)\right) W(y|x)}{\sum_{\mathcal{Y}} \exp\left(s \tilde{D}(x,y)\right) W(y|x)} ; \ s \ge 0 \nonumber \\
&= \frac{W(y|x)^{1-s} f(y)^s}{\sum_{\mathcal{Y}} W(y|x)^{1-s} f(y)^s} ; \ s \ge 0, \label{eq:V_equation}
\end{align}
where $\tilde{D}(x,y)$ is defined in \eqref{eq:D_def} and $s$ is chosen such that
\begin{equation}
\gamma'(s) = \sum_{\mathcal{X,Y}} P(x) V(y|x) \tilde{D}(x,y) = \frac{D_0}{n} .
\label{eq:D0_constraint}
\end{equation} 

Next, we derive an upper bound for $P_1$. If $\mathbf{u_0}$ is the transmitted sequence, $\mathbf{u'}$ is a randomly chosen sequence having the same subblock-composition as $\mathbf{u_0}$, $V_0$ is the set of sequences $\mathbf{v}$ for which $\tilde{D}(\mathbf{u_0,v}) \le D_0$, ${V'}_{\mathbf{v}}$ is the set of sequences $\mathbf{v'}$ that have the same subblock-composition as $\mathbf{v}$, ${V'}_{0\mathbf{v}}$ is the subset of ${V'}_{\mathbf{v}}$ for which $\tilde{D}(\mathbf{u_0,v'}) \le \tilde{D}(\mathbf{u_0,v})$, then:
\begin{align}
P_1 &= \sum_{V_0} W^n(\mathbf{v|u_0}) Pr[\tilde{D}(\mathbf{u',v}) \le \tilde{D}(\mathbf{u_0,v})] \nonumber \\
&\overset{(a)}= \sum_{V_0} W^n(\mathbf{v|u_0}) Pr[\tilde{D}(\mathbf{u_0,v'}) \le \tilde{D}(\mathbf{u_0,v})] \nonumber  \\ 
&= \sum_{V_0} W^n(\mathbf{v|u_0}) \ \frac{|{V'}_{0\mathbf{v}}|}{|{V'}_{\mathbf{v}}|}  \nonumber \\
&\overset{(b)}= \sum_{V_0} \ \frac{\displaystyle \sum_{{V'}_{0\mathbf{v}}} W^n(\mathbf{v|u_0}) F(\mathbf{v'})}{\displaystyle \sum_{{V'}_{\mathbf{v}}} F(\mathbf{v'})}
\label{eq:P1_form}
\end{align}
where $(a)$ follows from Lemma~\ref{lemma:ReplaceUWithV}, and $(b)$ follows because $F(\mathbf{v'})$ defined in \eqref{eq:CapitalF_Def} depends only on the composition of $\mathbf{v'}$ and is same for all sequences in $V'_{\mathbf{v}}$.

We will first bound the denominator in \eqref{eq:P1_form}. Let $f(y'|x)$ denote a conditional probability distribution with $f(y') = \sum_{\mathcal{X}} P(x) f(y'|x)$. This defines $F(\mathbf{v'|u})$ obtained as the product of the values of $f(y'|x)$ for each corresponding pair  of events of the sequences $\mathbf{v'}$ and $\mathbf{u}$. If $\mathbf{u}$ is \emph{any} sequence consisting of letters $x \in \mathcal{X}$, then we define $P(\mathbf{u}) = \prod_{\mathcal{X}} P(x)^{n(x)}$ where $n(x)$ is the number of occurrences of letter $x$ in $\mathbf{u}$. If $U$ is the space of all possible length $n$ sequences consisting of letters $x \in \mathcal{X}$, and $U_0$ denotes the subset of sequences having the given subblock-composition, then
\begin{equation}
F(\mathbf{v'}) = \sum_{U} P(\mathbf{u}) F(\mathbf{v'|u}) \ge \sum_{U_0} P(\mathbf{u}) F(\mathbf{v'|u})
\end{equation}
It follows that 
\begin{align}
\sum_{{V'}_{\mathbf{v}}} F(\mathbf{v'})\ &\ge \ \sum_{U_0} P(\mathbf{u}) \sum_{{V'}_{\mathbf{v}}} F(\mathbf{v'|u}) \\
&\overset{(c)}= \left( \sum_{U_0} P(\mathbf{u}) \right) \ \left( \sum_{{V'}_{\mathbf{v}}} F(\mathbf{v'|u}) \right) 
\end{align}
where $(c)$ follows because $\sum_{{V'}_{\mathbf{v}}} F(\mathbf{v'|u})$ is same for all $\mathbf{u} \in U_0$. Further,
\begin{align}
\sum_{U_0} P(\mathbf{u}) &= \left(\frac{L!}{\prod_{\mathcal{X}} (L\, P(x))!}\right)^m \ \left(\prod_{\mathcal{X}} P(x)^{L\, P(x)}\right)^m \nonumber \\ 
&= \exp\left(-n \, r(L,P)\right) 
\end{align}
with $r(L,P)$ given by \eqref{eq:RateLoss}. Thus, we have
\begin{equation}
\sum_{{V'}_{\mathbf{v}}} F(\mathbf{v'}) \ge \exp\left(-n r(L,P)\right) \sum_{{V'}_{\mathbf{v}}} F(\mathbf{v'|u}).
\label{eq:P1_DenomBound}
\end{equation} 

We will now bound the numerator in \eqref{eq:P1_form}. For each $x \in \mathcal{X}$, define the logarithm of the moment-generating function 
\begin{align}
\gamma_x(w_1,w_2) := \log & \sum_{y \in \mathcal{Y}} \sum_{y' \in \mathcal{Y}} \exp ( \, w_1 \tilde{D}(x,y) \ +  \nonumber \\
& w_2 [\tilde{D}(x,y') - \tilde{D}(x,y)] \, ) f(y') W(y|x)
\label{eq:Gammak_Def}
\end{align}
where $w_1$ and $w_2$ are parameters associated with the random variables $\tilde{D}(x,y)$ and $\tilde{D}(x,y')-\tilde{D}(x,y)$, respectively. Define the tilted probability distributions
\begin{align}
Q_0(y,y'|x) := \exp(\, & w_1 \tilde{D}(x,y) + w_2 [\tilde{D}(x,y') - \tilde{D}(x,y)] \nonumber \\
&- \gamma_x(w_1,w_2) \, ) \, f(y') W(y|x) 
\end{align}
\begin{align}
Q_0(\mathbf{v,v'|u_0}) := \exp(\, & w_1 \tilde{D}(\mathbf{u_0,v}) + w_2 [\tilde{D}(\mathbf{u_0,v'}) - \tilde{D}(\mathbf{u_0,v})] \nonumber \\
&- n \gamma_0(w_1,w_2) \, ) F(\mathbf{v'}) W^n(\mathbf{v|u_0}) \label{eq:Q0Seq_TiltedProb}
\end{align}
where $\gamma_0(w_1,w_2) = \sum_{\mathcal{X}} P(x) \gamma_x(w_1,w_2)$. From \eqref{eq:Q0Seq_TiltedProb}, it follows that
\begin{align}
\sum_{{V'}_{0\mathbf{v}}} W^n(\mathbf{v|u_0}) F(\mathbf{v'})&\le \exp\left( n \gamma_0(w_1,w_2) - w_1 D(\mathbf{u_0,v}) \right) \nonumber \\
& \times \sum_{{V'}_{0\mathbf{v}}} Q_0(\mathbf{v,v'|u_0})  ; \ w_2 \le 0
\label{eq:P1_Num}
\end{align}
The RHS of \eqref{eq:P1_Num} is minimized by setting $w_1 = 2 w_2+1$, in which case we have
\begin{align}
Q_0(y,y'|x) &= Q_0(y|x) Q_0(y'|x)
\label{eq:Q0_symmetry} \\
Q_0(y|x) &= \frac{W(y|x)^{(1-w_1)/2} \ f(y)^{(1+w_1)/2}}{ \sum_{\mathcal{Y}} W(y|x)^{(1-w_1)/2} \ f(y)^{(1+w_1)/2}} 
\label{eq:Q0_symmetry2} \\
Q_0(\mathbf{v,v'|u_0}) &= Q_0(\mathbf{v|u_0}) Q_0(\mathbf{v'|u_0}) \label{eq:Q0_symmetry3}
\end{align}
Using \eqref{eq:P1_Num}, \eqref{eq:Q0_symmetry3}, and the fact that ${V'}_{0\mathbf{v}} \subset {V'}_{\mathbf{v}}$, we have
\begin{align}
\sum_{{V'}_{0\mathbf{v}}} W^n(\mathbf{v|u_0})  F(\mathbf{v'})& \le \exp\left(n \gamma_0(w_1,w_2) - w_1 \tilde{D}(\mathbf{u_0,v})\right)  \nonumber \\
& \times Q_0(\mathbf{v|u_0}) \sum_{{V'}_{\mathbf{v}}} Q_0(\mathbf{v'|u_0}) ; \ w_1 \le 1
\label{eq:P1_Num_UB}
\end{align}
If we let $f(y'|x) = Q_0(y'|x)$ then $F(\mathbf{v'}) = Q_0(\mathbf{v'}|\mathbf{u_0})$ and $P_1$ can be bounded using \eqref{eq:P1_form}, \eqref{eq:P1_DenomBound}, and \eqref{eq:P1_Num_UB} as

\begin{align}
P_1 \le \exp &\left( n r(L,P)+n\gamma_0(w_1,w_2) \right) \times  \nonumber \\
& \sum_{V_0} \exp\left(-w_1 \tilde{D}(\mathbf{u_0,v}) \right) Q_0(\mathbf{v|u_0}) ; \ w_1 \le 1 \nonumber
\label{eq:P1_bound3}
\end{align}
Since all the sequence $\mathbf{v}$ belonging to $V_0$ have a distance from $\mathbf{u_0}$ which does not exceed $D_0$, we have
\begin{equation}
P_1 \le \exp \left( n r(L,P) + n\gamma_0(w_1,w_2) - w_1 D_0 \right) ; \ w_1 \le 0
\label{eq:P1_bound4}
\end{equation}
From \eqref{eq:D0_constraint}, we know that $D_0 = n \gamma'(s)$, and RHS of \eqref{eq:P1_bound4} is minimized when $w_1 = 2s-1, w_2 = s-1$, and $s$ satisfies  $0 \le s \le 1/2$. In this case, we have
\begin{align}
Q_0(y|x) &= V(y|x) \label{eq:Q0_Equal_V} ,\\
\gamma_0(w_1,w_2) &= 2 \gamma (s) \ ,
\end{align}
where $V(y|x)$ is given by \eqref{eq:V_equation}. Now \eqref{eq:P1_bound4} takes the form
\begin{equation}
P_1 \le \exp\left(-n [(2s-1)\gamma'(s) - 2 \gamma(s) - r(L,P)] \right)  ; \ 0 \le s \le 1/2
\label{eq:P1_bound}
\end{equation}
Note that since $f(y|x) = Q_0(y|x)$, it follows from \eqref{eq:Q0_Equal_V} that 
\begin{equation}
f(y) = PV(y) \ .
\label{eq:f_Equal_PV}
\end{equation}
Since $R = \frac{\log M}{n}$, and $P_e < M P_1 + P_2$, we have
\begin{align}
P_e < &\exp\left( -n [(2s-1)\gamma'(s) - 2 \gamma(s) - (R+r(L,P))]\right) \nonumber \\
&+ \exp\left( -n [s\gamma'(s) - \gamma(s)] \right) ; \ 0 \le s \le 1/2 .
\label{eq:Pe_Bound1}
\end{align}
Now, if we choose $s$ such that
\begin{equation}
(s-1)\gamma'(s) - \gamma(s) = R + r(L,P) \ ; \ 0 \le s \le 1/2 \ ,
\label{eq:Fixing_s}
\end{equation} 
then it follows from \eqref{eq:Pe_Bound1}, \eqref{eq:Fixing_s}, and \eqref{eq:Beta_Def} that
\begin{equation}
P_e < 2\exp\left( -n D(V||W |P)\right) .
\label{eq:Pe_Bound2}
\end{equation}
From \eqref{eq:V_equation} and \eqref{eq:f_Equal_PV} we have
\begin{equation}
V(y|x) = \frac{W(y|x)^{1-s} PV(y)^s}{\sum_y W(y|x)^{1-s} PV(y)^s} .
\label{eq:V_equation2}
\end{equation}
Now $(s-1)\gamma'(s) - \gamma(s)$ is a decreasing function of $s$ for $0 \le s \le 1/2$ (its derivative is $(s-1)\gamma''(s)$). Let $\hat{R}$ denote its value at $s=1/2$ (that is $\hat{R} = -0.5\gamma'(0.5)-\gamma(0.5)$. From \eqref{eq:Beta_Def}, \eqref{eq:D0_constraint}, and \eqref{eq:f_Equal_PV}, it follows that condition \eqref{eq:Fixing_s} can be equivalently be expressed as 
\begin{equation}
I(P,V) = R + r(L,P) \ , \mbox{\ if } R + r(L,P) \ge \hat{R}.
\label{eq:Constrain_IPV}
\end{equation}
If we let
\begin{equation}
R' = R + r(L,P) \ ,
\label{eq:Rprime}
\end{equation}
then using Lemma~\ref{lemma:StructureOfV}, \eqref{eq:V_equation2}, and \eqref{eq:Constrain_IPV}, we observe that $D(V||W|P) = E_{sp}\left(R',P,W\right)$, and \eqref{eq:Pe_Bound2} is equivalent to
\begin{equation}
P_e < 2 \exp\left(-n E_{sp}\left(R',P,W\right)\right) ;  \ \ \mathrm{if \ } R' \ge \hat{R} .
\label{eq:AvProbError_v1}
\end{equation}
Note that for $0 \le s \le 1/2$, we have $E_{sp}\left(R',P,W\right) = s\gamma'(s) - \gamma(s)$ and $R' = (s-1)\gamma'(s) - \gamma(s)$. Thus
\begin{equation}
\frac{d }{d R'}E_{sp}\left(R',P,W\right) = -\frac{s}{1-s} \ ; \ \ 0 \le s \le 1/2
\end{equation}
and the slope of $E_{sp}\left(R',P,W\right)$ with respect to $R'$ at $s=1/2$ (corresponding to $R' = \hat{R}$) is $-1$.

For obtaining a bound on $P_e$ when $R' < \hat{R}$, we let radius $D_0 = \infty$. In this case,
\begin{equation}
P_1 = \sum_{\tilde{V}} P(\mathbf{v|u_0}) \ Pr[\tilde{D}(\mathbf{u',v}) \le \tilde{D}(\mathbf{u_0,v})] 
\label{eq:P1_equation_v2}
\end{equation}
where $\tilde{V}$ is the space of all possible output sequences. Note that $P_2 = 0$ when $D_0 = \infty$. The upper bound for $P_1$ in \eqref{eq:P1_equation_v2} can be calculated the same way as before. However, the condition $D(\mathbf{u_0,v}) \le D_0$ is eliminated by setting $w_1 = 0$ in \eqref{eq:Gammak_Def}. Here, the upper bound for $P_1$ is obtained by setting $w_2 = s-1$ and $s=1/2$. The probability of error in this case is bounded as 
\begin{align}
P_e &< \exp\left(-n \left(-2\gamma(0.5) - R'\right)\right) ; \mbox{\ if } R' < \hat{R} \nonumber \\
&= \exp\left(-n \left(E_{sp}(\hat{R},P,W) + \hat{R}- R'\right)\right) ; \mbox{\ if } R' < \hat{R} 
\label{eq:AvProbError_v2}
\end{align}
and the proof is complete by combining \eqref{eq:Rprime}, \eqref{eq:AvProbError_v1} and \eqref{eq:AvProbError_v2}. 

\section{Proof of Theorem~\ref{th:SECC_NotUniform}}
\label{app:SECC_NotUniform}
For the proof, we will construct a simple example of a symmetric DMC for which the uniform distribution over $\mathcal{A}$ does not achieve SECC capacity.
\begin{figure}[h]
\centering
\includegraphics[width=\myfigwidth\textwidth]{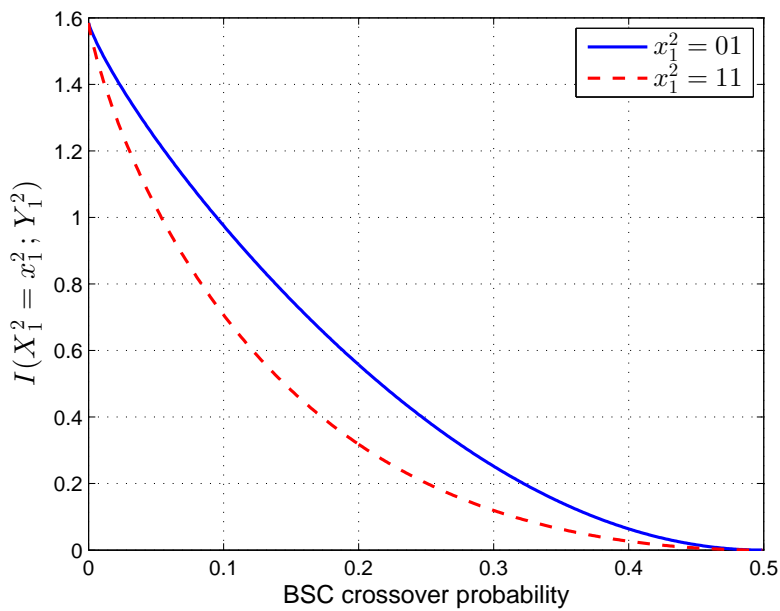}
\caption{$I(X_1^L=x_1^L ; Y_1^L)$ versus BSC crossover probability $p_0$ for $L=2$.}
\label{Fig:SECC_NotSymmetric}
\end{figure}

Consider the following parameters for a BSC with crossover probability $p_0$:
\begin{equation}
b(0)=0, \ b(1)=1, \ B=0.5, \ L=2, \ 0 < p_0 < 0.5 \ .
\label{eq:BSC_ParameterList1}
\end{equation}
With the above parameters, the input alphabet for the induced vector channel is given by $\mathcal{A} = \{01, 10, 11\}$. A uniform distribution over $\mathcal{A}$ will achieve SECC capacity if and only if $I(X_1^L=x_1^L ; Y_1^L )$ is same for all $x_1^L \in \mathcal{A}$ \cite[Thm.~4.5.1]{GallagerBook68}, where
\begin{equation}
I(X_1^L=x_1^L ; Y_1^L ) = \sum_{y_1^L \in \mathcal{Y}^L} W^L(y_1^L | x_1^L) \log \frac{ |\mathcal{A}| \ W^L(y_1^L | x_1^L)}{\displaystyle \sum_{\tilde{x}_1^L \in \mathcal{A}} W^L(y_1^L | \tilde{x}_1^L)} .
\end{equation}
The proof is completed by numerically verifying that for BSC having parameters given by \eqref{eq:BSC_ParameterList1}, $I(X_1^L=01 ; Y_1^L) \neq I(X_1^L=11 ; Y_1^L)$. Fig.~\ref{Fig:SECC_NotSymmetric} shows that $I(X_1^L=01 ; Y_1^L)$ and $I(X_1^L=11 ; Y_1^L)$ are different when $0 < p_0 < 0.5$.  \hfill $\blacksquare$

\balance
\bibliographystyle{IEEEtran}
\bibliography{abrv,conf_abrv,mybibfile}

\end{document}